\documentclass[12pt,preprint,twocolumn,twocolappendix]{aastex701}

\newcommand{\Hii}{\ion{H}{0ii}\,}

\submitjournal{ApJ}

\usepackage{graphicx} 
\usepackage{xcolor} 
\usepackage{amsmath}	

\defcitealias{frew16}{F16} 
\defcitealias{chornay21}{CW21} 
\defcitealias{bucciarelli23}{BS23} 
\defcitealias{deng26}{D26}

\providecommand{\parallax}{\ensuremath{\varpi}} 

\providecommand{\parallaxsd}{\ensuremath{\sigma_\parallax}}

\begin{document}

\title{Recalibration of the H$\alpha$ surface brightness--radius relation for planetary nebulae using Gaia DR3: new
distances and the Milky Way oxygen radial gradient}



\author[orcid=0000-0002-7103-8036]{Oscar Cavichia} \affiliation{Instituto de F{\'i}sica e Qu{\'i}mica, Universidade
Federal de Itajub{\'a}\\ Av. BPS, 1303, 37500-903, Itajub{\'a}-MG, Brazil } \email[show]{cavichia@unifei.edu.br}

\author[0000-0002-0596-9115]{Hektor Monteiro} \affiliation{Instituto de F{\'i}sica e Qu{\'i}mica, Universidade Federal
de Itajub{\'a}\\ Av. BPS, 1303, 37500-903, Itajub{\'a}-MG, Brazil } \email[]{hmonteiro@unifei.edu.br}

\author[0000-0001-8009-231X]{Miguel Cervi\~{n}o} \affiliation{Centro de Astrobiolog\'ia (CSIC/INTA), ESAC Campus\\
Camino Bajo del Castillo s/n, E-28692 Villanueva de la Ca\~{n}ada, Spain } \email[]{mcs@cab.inta-csic.es}

\author{Adalberto R. da Cunha-Silva} \affiliation{Instituto de F{\'i}sica e Qu{\'i}mica, Universidade Federal de
Itajub{\'a}\\ Av. BPS, 1303, 37500-903, Itajub{\'a}-MG, Brazil } \email[]{adalbertocunhasilva@gmail.com}

\author[0000-0002-9823-3296]{Walter J. Maciel} \affiliation{Instituto de Astronomia, Geof{\'i}sica e Ci\^{e}ncias
Atmosf\'{e}ricas, Universidade de S\~{a}o Paulo\\ Rua do Mat\~{a}o, 1226, 05508-090, S\~{a}o Paulo-SP, Brazil }
\email[]{wjmaciel@iag.usp.br }

\author[0000-0003-1097-3247]{Andr\'{e} F. S. Cardoso} \affiliation{Instituto de F{\'i}sica e Qu{\'i}mica, Universidade
Federal de Itajub{\'a}\\ Av. BPS, 1303, 37500-903, Itajub{\'a}-MG, Brazil } \email[]{afelipe2992@unifei.edu.br}
\affiliation{N{\'u}cleo Cosmo-Ufes \& Departamento de F{\'i}sica -- Universidade Federal do Esp{\'i}rito Santo\\
29075-910, Vit{\'o}ria, ES, Brazil }

\begin{abstract}


The spatial distribution of chemical elements in the Galactic disk provides key constraints on models of galaxy
evolution. However, studies using planetary nebulae (PNe) as tracers have been historically limited by large
uncertainties in their distances. To overcome the long-standing distance uncertainties, we recalibrated the H$\alpha$
surface brightness--radius relation from Frew et al. with \textit{Gaia} DR3 parallaxes, deriving distances for 1,130 PNe
of which 415 have Bayesian distances based on  \textit{Gaia} DR3 parallaxes. The O/H radial gradient for 231 disk PNe is
fitted considering three models: a single linear gradient and segmented linear fits with one or two breaks. The
segmented fits indicate a change in slope near the solar radius ($R \sim 8$ kpc), with a flatter or slightly positive
gradient inward and a steeper negative gradient outward. This feature may reflect changes in star formation efficiency
driven by the Galactic bar or the corotation resonance of the spiral arms. The breaks in the metallicity radial
gradients observed in this work may result from the  superposition of distinct stellar populations associated with the
thin and thick disks. The two-dimensional O/H distribution in the Galactic plane supports the adopted distances and
reveals modest azimuthal asymmetry, with enhanced abundances near the bar at positive longitudes, and a bimodal
abundance structure between the inner and outer solar regions. Our results provide new constraints on the chemical
evolution of the Milky Way, the impact of non-axisymmetric structures, and the possible existence of distinct radial
abundance regimes across the Galactic disk.

\end{abstract}




\section{Introduction \label{sec:intro}}

The absolute amount of metals, the relative abundance of different elements, and their spatial distributions in a galaxy
directly depend on the galactic evolutionary history and, therefore, serve as crucial constraints for chemical evolution
models \citep{gibson13}. In this context, the radial gradient of chemical abundances in the disks of spiral galaxies is
one of the most important constraints for galaxy chemical evolution models \citep{henry99}. The gradient was first
observed in the Milky Way by \citet{shaver83}, revealing a radial decrease in oxygen abundances, with higher values in
the central regions. Subsequently, the abundance gradient was observed in other external galaxies, as shown in
\citet{mccall85}, \citet{zaritsky94}, and \citet{vanzee98}, and is now well-established in the local universe
\citep{sanchez14}. Currently, a typical gradient of $\sim-0.05$ dex/kpc is found for the Milky Way disk
\citep{molla19a}. This gradient results from various physical processes acting from galaxy formation to the present,
including gas infall/outflow, stellar formation history, initial mass function, and radial gas flows. \citet{gibson13}
demonstrated that the existence and evolution of these gradients strongly depend on the prescriptions for star formation
and gas infall included in the simulations. While recognizing the concurrence of observations for a negative gradient in
the radial interval of Galactocentric distances $4 \lesssim R \lesssim 10$ kpc in the Milky Way disk and 0.5 to
2\textit{R}\textsubscript{e} in other spiral galaxies, where \textit{R}\textsubscript{e} is the effective radius
(half-light radius), there are evidences of a flattening of the gradient in spiral galaxies beyond
2\textit{R}\textsubscript{e} \citep{sanchez14,sanchez-menguiano16}. The advent of multi-object and integral field
spectroscopy has ushered in instruments with expansive fields of view, enabling a new generation of surveys focused on
the observation of \Hii regions in external galaxies. Surveys like CALIFA \citep{sanchez12} and MANGA \citep{bundy15}
have extensively observed hundreds of \Hii regions in the disks of nearby spiral galaxies, providing comprehensive 2D
coverage. The results presented by \citet{sanchez14}, based on observations of over 7000 \Hii regions in 306 spiral
galaxies, indicate evidence of a flattening of the O/H gradient beyond 2\textit{R}\textsubscript{e}, consistent with
earlier studies focusing primarily on a few objects \citep[e.g.,][]{bresolin09}. These findings were further supported
by \citet{sanchez-menguiano16}.

In the case of the Milky Way, a significant debate exists in the literature regarding the constancy of the gradient
across different Galactocentric distances. In the outer regions of the Milky Way ($R >$ 10 kpc) a flattening of the
radial gradient of the disk is noted by data from different tracers such as open clusters
\citep{lepine11,monteiro21,magrini23}, \Hii regions \citep{esteban13}, Cepheid stars \citep{genovali14, minniti20}, and
planetary nebulae \citep[PNe,][]{maciel09,stanghellini18}. These studies indicate that abundances remain relatively
constant in the outer disk as $R$ increases. Subsequently, however, Esteban et al. found no evidence for a slope
change in the outer disk after incorporating new high-quality data into their dataset \citep{esteban17,
arellano-cordova20, mendez-delgado22, martinez-hernandez26}. Similarly, other works suggest that the gradient maintains
a consistent slope across the entire optical disk, as evidenced by e.g. \citet{balser11, fernandez-martin17, wenger19}
for \Hii regions, as well as e.g. \citet{stanghellini10,pagomenos18} and \citet[][hereafter BS23]{bucciarelli23} for
PNe. Conversely, \citet{henry10} analyzed disk PNe data to derive the radial gradient and find evidence suggesting a
steepening of the gradient at large Galactocentric distances, though they emphasized the need for additional data to
confirm this trend.

Due to discrepancies within the literature, establishing whether the gradient maintains a consistent slope across all
observed radii in the Milky Way proves challenging. Part of the disparity between different studies stems from
uncertainties in determining chemical abundances and/or distances. Estimating reliable distances for Galactic PNe proves
particularly challenging, as no single physical parameter directly depends on the distance. Statistical distance methods
have therefore been developed to derive PNe distances \citep[see][hereafter F16, for a comprehensive review]{frew16}.
The statistical distance scale from \citetalias{frew16}, based on an empirical relation between $\mathrm{H}\alpha$
surface brightness ($S_{\mathrm{H}\alpha}$) and the intrinsic radius of the PNe, has made an important improvement in
the determination of PNe distances. This relation was calibrated using data for 322 PNe, of which 206 are Galactic and
126 are extragalactic objects. They provide three relations, one for the full sample, one for optically thick objects,
and one for optically thin PNe. Distances for 1,133 PNe are given, 515 of them being classified as optically thick or
optically thin nebulae.

However, statistical distances are subject to considerable uncertainties, often due to factors of two or more, as
discussed by \citetalias{frew16}. Consequently, trigonometric techniques remain the direct and reliable individual
method to determine the distances of the PNe when the parallax errors are small \citep[parallax uncertainty over
parallax, $\parallaxsd/\parallax$, less than 0.15,][]{hernandez-juarez24}. In this regard, the recent \textit{Gaia}
mission \citep{gaia16,gaia18} has revolutionized the determination of the distances for the Galactic PNe
\citep{gonzalez-santamria21,chornay21}. Before \textit{Gaia}, only a dozen of PNe located very close to the Sun had
distances estimated from parallaxes \citepalias{frew16}. \textit{Gaia} has significantly altered this by providing
precise parallaxes for hundreds of Galactic PNe. The identification of PNe in \textit{Gaia} archive has been improved in
the last few years. \citet{stanghellini17} identified 8 central stars of PNe (CSPNe) in the DR1. \citet{kimeswenger18}
have manually identified 382 \textit{Gaia} sources in DR2 that match the PNe coordinates. \citet{stanghellini20} matched
the astrometry of CSPNe with DR2 finding 430 sources. \citet{ali22} matched \textit{Gaia} sources with the HASH catalog
\citep{parker06} and \citet{weidmann20} CSPNe coordinates finding 603 sources in common. \citet{gonzalez-santamria21}
and \citet{chornay20,chornay21} developed more sophisticated methods based on the color and geometric distance of the
\textit{Gaia} sources in EDR3 to identify 2035 and 2117 PNe correspondences, respectively.

The \textit{Gaia} DR3 \citep{gaia16,gaia_dr3_23} provides unprecedented astrometric parameters, complemented by
photometric and spectroscopic data, for hundreds of CSPNe. Even after \textit{Gaia} observations, statistical distances
for PNe will be continuously used, since many PNe will not have \textit{Gaia} distance estimates because of the CSPNe
are not well identified or in the cases where the parallaxes have a considerable error.  In this regard, statistical
distance methods for PNe have been recalibrated by using \textit{Gaia} astrometric data, as demonstrated by
\citet{stanghellini20,ali22} and \citetalias{bucciarelli23}.

In this paper, our aim is to revisit the issue of PNe abundance gradients relying on reliable distances determined
directly from astrometric parallaxes provided by Gaia data and from statistical distance scale. We present a
recalibration of the PN distance scale from \citetalias{frew16} using the best parallaxes available from \textit{Gaia}
DR3 \citep{lindegren21}, enhancing its ability as a robust distance indicator for the many PNe which will not have
\textit{Gaia} distance estimates.

This work is organized as follows: in Section \ref{sec:data} the source selection from the \textit{Gaia} DR3 data, the
adopted chemical abundances, the adopted methodology to derive the PNe distances, and the criterion to separate the thin
and thick disk samples are presented. In Section \ref{sec:results} the main results are described related with the
radial and azimuthal oxygen abundances distribution and in Section \ref{sec:conclusions} the discussion and conclusions
are presented.

\section{Methods} \label{sec:data}

\subsection{Gaia DR3 source identification}

The \textit{Gaia} space mission was launched and operated by the European Space Agency (ESA) to provide a detailed
three-dimensional map of the Milky Way Galaxy. The \textit{Gaia} Data Release 3 (\textit{Gaia} DR3) was
published in June 2022 and provides full astrometric solution — positions on the sky ($\alpha$, $\delta$), parallaxes
(\parallax), and proper motions ($\mu$) — for around 1.46 billion sources, $G$ magnitudes for around 1.806 billion
sources, $G_{\textrm{BP}}$ and $G_{\textrm{RP}}$ magnitudes for around 1.54 billion and 1.55 billion sources,
respectively. Compared to \textit{Gaia} DR2, the new release offers significant improvements in astrometric and
photometric accuracy, precision, and homogeneity \citep{gaia_dr3_23}.

 The identification of \textit{Gaia} sources that match the CSPNe has demonstrated to be challenge, as the CSPNe are
 faint and sometimes undetectable, or may have multiple candidate sources within the central region of the nebula.
 \citet{gonzalez-santamria21} developed a method to identify CSPNe in \textit{Gaia} EDR3  based on proximity to the
 geometric center of the nebula and photometric color. Similarly, \citet{chornay20} and \citet[][hereafter
 CW21]{chornay21} employed a likelihood ratio method to cross-match known PNe with \textit{Gaia} EDR3 sources, using
 empirically derived positional and color distributions to assess candidate likelihoods. We adopted a procedure in this
 paper similar to \citet{gonzalez-santamria21} but with some enhancements, as detailed below.

We began by constructing a catalog of PNe coordinates. Whenever available, CSPNe coordinates were taken from
\citet{weidmann20}; otherwise, we used our internal PNe database \citep{maciel15} and coordinates from \citet{frew13},
\citetalias{frew16}, and \citetalias{chornay21}. Angular diameters were preferentially adopted from \citetalias{frew16},
supplemented by data from \citet{tylenda03}, \citet{stanghellini08}, and \citet{acker92}. The $\mathrm{H}\alpha$ surface
brightness are obtained from \citetalias{frew16}. This compilation resulted in a sample of 1,200 PNe with reliable
coordinates and angular diameters, of which 1,130 PNe have reliable H$\alpha$ fluxes.

We queried the \textit{Gaia} archive around the compiled coordinates, setting the search radius from half the nebular
radius up to a maximum of 80\% of the radius. For cases where half the nebular radius was less than $1.8''$, we used
this value as the minimum radius, following the \textit{Gaia} documentation formula: $1'' + \Delta_{\rm epoch}, [{\rm
yr}] \times 0.050''/{\rm yr} = 1.8''$, where $\Delta_{\rm epoch} = 2016 - 2000 = 16$ yr and $0.050''/{\rm yr}$
corresponds to the maximum expected proper motion for 99.8\% of \textit{Gaia} sources. In addition, it is essential to
assess the quality of the astrometric solutions. This is commonly evaluated using the Renormalized Unit Weight Error
(RUWE) parameter, which quantifies the goodness-of-fit of the astrometric model to the observations \citep{lindegren18}.
RUWE values greater than 1.4 typically indicate unreliable or problematic astrometric solutions.\footnote{See the
\textit{Gaia} technical note GAIA-C3-TN-LU-LL-124-01 and documentation at
\url{https://www.cosmos.esa.int/web/gaia/dr3}.} The parameter \texttt{visibility\_periods\_used} is the number
of distinct observation epochs (visibility periods) used in the solution \citep{lindegren21}. In this work, we exclude
parallaxes from sources with RUWE $> 1.4$ and applied the condition \texttt{visibility\_periods\_used} $> 8$ to ensure
reliable astrometric solutions. This combination removes most binary systems, variable stars, and poorly
measured sources \citep{lindegren18}. The search yielded 27,401 \textit{Gaia} sources for the 1,200 PNe in our
database. As an additional and necessary step, parallaxes were corrected for the zero-point offset following the
prescription of \citet{lindegren21}.

To identify the most probable CSPNe, we implemented the color-distance method of \citet{gonzalez-santamria21}. We
assumed a minimum CSPN temperature of 13,000 K for the post-AGB transition phase \citep{weidmann20}, and applied the
color--temperature relation from \citet{jordi10}, where $G_{\rm BP} - G_{\rm RP} = -0.2$ corresponds to $T_{\rm eff} =
14,477$ K, establishing this value as the upper color limit for candidate selection. For sources lacking $G_{\rm BP} -
G_{\rm RP}$ colors but with available $T_{\rm eff}$ from DR3, we used the inverse relation from \citet{jordi10} to
estimate the corresponding color.

To further refine the selection, we used the \texttt{astropy} package \citep{astropy22} to account for proper motion of
the \textit{Gaia} sources and calculate coordinates at the reference epoch J2000. Interstellar extinction corrections
were applied to the photometry using the Bayestar \citep{green19}, Marshall \citep{marshall06}, and SFD
\citep{schlegel98} extinction maps, in that order of priority. Visual extinctions were converted to \textit{Gaia} band
extinctions using the relations provided by \citet{gentile-fusillo21}.

We adopted the classification scheme from \citet{gonzalez-santamria21}, grouping the identifications into three
reliability classes: A (most reliable), B, and C (least reliable). In our sample, 48\% of the CSPNe fell into group A,
36\% into group B, and 16\% into group C. This represents a notable improvement over \citet{gonzalez-santamria21}, who
report 31.6\%, 30.9\%, and 37.5\% in groups A, B, and C, respectively.

 \begin{figure}[ht!] \includegraphics[width = 1.0\columnwidth]{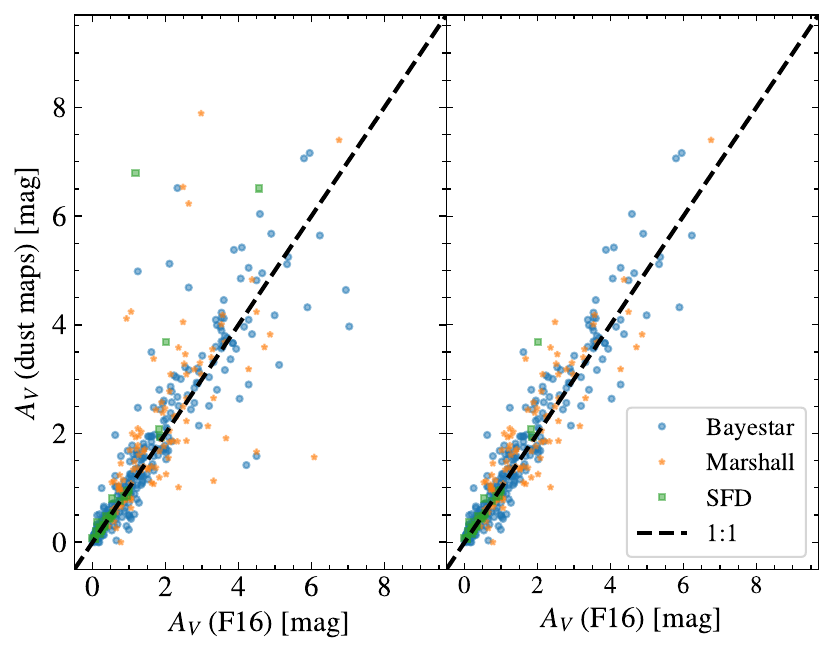} \caption{Left: extinction in the visual band
 obtained in this work from the Bayestar \citep{green19}, Marshall \citep{marshall06} and SFD \citep{schlegel98} dust
 maps as a function of the extinction obtained from \citetalias{frew16}, as labeled. Right: the same as the left panel
 but excluding PNe that are $3 \sigma$ from the identity line. \label{fig:extinction}} \end{figure}

We then cross-matched our high-confidence identifications with those from \citetalias{chornay21}, selecting their
``true" PNe with reliability scores $> 0.8$. This yielded 462 PNe in common. As an additional consistency check, we
compared the visual extinction values derived for each PN from the Balmer decrement reported in \citetalias{frew16} with
the extinction from Galactic dust maps, as shown in the left panel of Fig.\ref{fig:extinction}. While the agreement is
generally good, some outliers deviate significantly from the identity line, suggesting either incorrect CSPN
associations, non-standard extinction values, or PNe with high internal dust content. We excluded sources lying more
than $3\sigma$ from the identity relation, as illustrated in the right panel of Fig.\ref{fig:extinction}. Our final
sample consists of 436 PNe with a high probability of correct association between the \textit{Gaia} sources and their
corresponding CSPNe. From this total, 415 PNe have H$\alpha$ fluxes necessary to calculate the distances (see Section
\ref{sec:bayes}). Table \ref{tab:gaia_id} lists the results for the \textit{Gaia} DR3 source identification and the
obtained parameters from DR3. The description of the columns is given as a table note.

\begin{deluxetable*}{llccccccccccc}
\tabletypesize{\scriptsize}
\tablewidth{0pt}
\setlength{\tabcolsep}{1.7pt}
\tablecaption{\textit{Gaia} DR3 source identification.\label{tab:gaia_id}}
\tablehead{
\colhead{PNG} & \colhead{Name} & \colhead{Source ID} & \colhead{RA$_{\mathrm{DR3}}$} & \colhead{DE$_{\mathrm{DR3}}$} & \colhead{$G$} & \colhead{$G_{\mathrm{BP}}-G_{\mathrm{RP}}$} & \colhead{RUWE }& \colhead{\parallax} & \colhead{Ang. sep.}  & \colhead{Case} & \colhead{$A_{\mathrm{V}}$} &  \colhead{$(G_{\mathrm{BP}}-G_{\mathrm{RP}})_0$}\\
\colhead{}   & \colhead{}  & \colhead{}  & \colhead{(\degr)} & \colhead{(\degr)} & \colhead{(mag)} & \colhead{(mag)} & \colhead{} &  \colhead{(mas)} & \colhead{(arcsec)}  & \colhead{}   & \colhead{(mag)} & \colhead{(mag)}\\
\colhead{(a)} & \colhead{(b)} & \colhead{(c)} & \colhead{(d)} & \colhead{(e)} & \colhead{(f)} & \colhead{(g)} & \colhead{(h)} & \colhead{(i)} & \colhead{(j)} & \colhead{(k)} & \colhead{(l)} & \colhead{(m)}
}
\startdata
000.0-06.8 & H 1-62 & 4045771305065496832 & 273.325 & -32.329 & 14.36 & 0.77 & 0.78 & $0.090 \pm 0.034$ & 0.77 & B & 1.36 & 0.10 \\
000.3+12.2 & IC 4634 & 4126115570219432448 & 255.390 & -21.826 & 13.85 & -0.15 & 0.62 & $0.391 \pm 0.043$ & 0.33 & A & 1.02 & -0.65 \\
000.4+04.4 & K 5-1 & 4109691718340733568 & 262.468 & -26.187 & 19.18 & 1.76 & 1.04 & $-0.149 \pm 0.395$ & 0.63 & A & 5.38 & -0.87 \\
000.6-01.3 & Bl 3-15 & 4056579538679085952 & 268.150 & -29.111 & 19.24 & 2.03 & 1.21 & $-0.255 \pm 0.474$ & 0.16 & A & 6.04 & -0.92 \\
000.9-04.8 & M 3-23 & 4049925328633027712 & 271.776 & -30.571 & 19.22 & 0.65 & 1.00 & $0.266 \pm 0.383$ & 0.70 & B & 2.31 & -0.48 \\
\enddata
\tablecomments{Table \ref{tab:gaia_id} is published in its entirety in the machine-readable format and available electronically at the CDS. A portion is shown here for guidance regarding its form and content. The columns of the table correspond to the following identifications:  (a) PN PNG number; (b) PN name; (c) \textit{Gaia} DR3 source identification (d) \textit{Gaia} right ascension coordinate in degrees; (e) \textit{Gaia} declination coordinate in degrees; (f) \textit{Gaia} $G$ magnitude; (g) \textit{Gaia} color; (h) Gaia RUWE parameter; (i) \textit{Gaia} corrected parallax; (j)
\textit{Gaia} source to CSPN angular separation; (k) identification case; (l) Galactic dust maps extinction; (m) \textit{Gaia} de-reddened color.}
\end{deluxetable*}

\subsection{Distances} \label{sec:distances}

The H$\alpha$ surface brightness--radius relation was used by \citetalias{frew16} to obtain statistical distances for
Galactic PNe. The scale was calibrated based on a large number of PNe from our Galaxy as well as other nearby galaxies.
To perform the calibration, the distances were determined by several different methods, including a small number of 13
PNe with precise trigonometric parallaxes and also other less reliable distances, including photometric distances,
eclipsing binaries, expansion parallaxes, physical membership of a PN in an open or globular star cluster, Galactic
bulge or a galaxy, among others \citepalias[see][for more details]{frew16}. \textit{Gaia} DR3 data allow a recalibration
of the H$\alpha$ surface brightness--radius relation obtained by \citetalias{frew16}, as a large number of PNe with
accurate astrometric parallaxes are available.

\subsubsection{Recalibration of the H$\alpha$ surface brightness--radius relation}

PNe with accurate parallaxes can be used to recalibrate the H$\alpha$ surface brightness--radius relation obtained by
\citetalias{frew16}. In the cases where the fractional parallax uncertainties are low, it is possible to calculate the
physical radius of the PN using $R_{\rm pc} = \theta/(206265\times \parallax)$, where $\theta$ is the PN angular radius
in arcsec. The intrinsic H$\alpha$ surface brightness in units of ${\rm erg}\,{\rm cm}^{-2} \, {\rm s}^{-1}\, {\rm
sr}^{-1}$ is calculated using the formula $S_{{\rm H}\alpha} = F_{{\rm H}\alpha}/(4\pi \theta^2)$, being that $F_{{\rm
H}\alpha}$ the H$\alpha$ flux and $\theta$ the angular diameter, both obtained from \citetalias{frew16}.

\begin{figure}[ht!] \centering \includegraphics[width = 1.0\columnwidth]{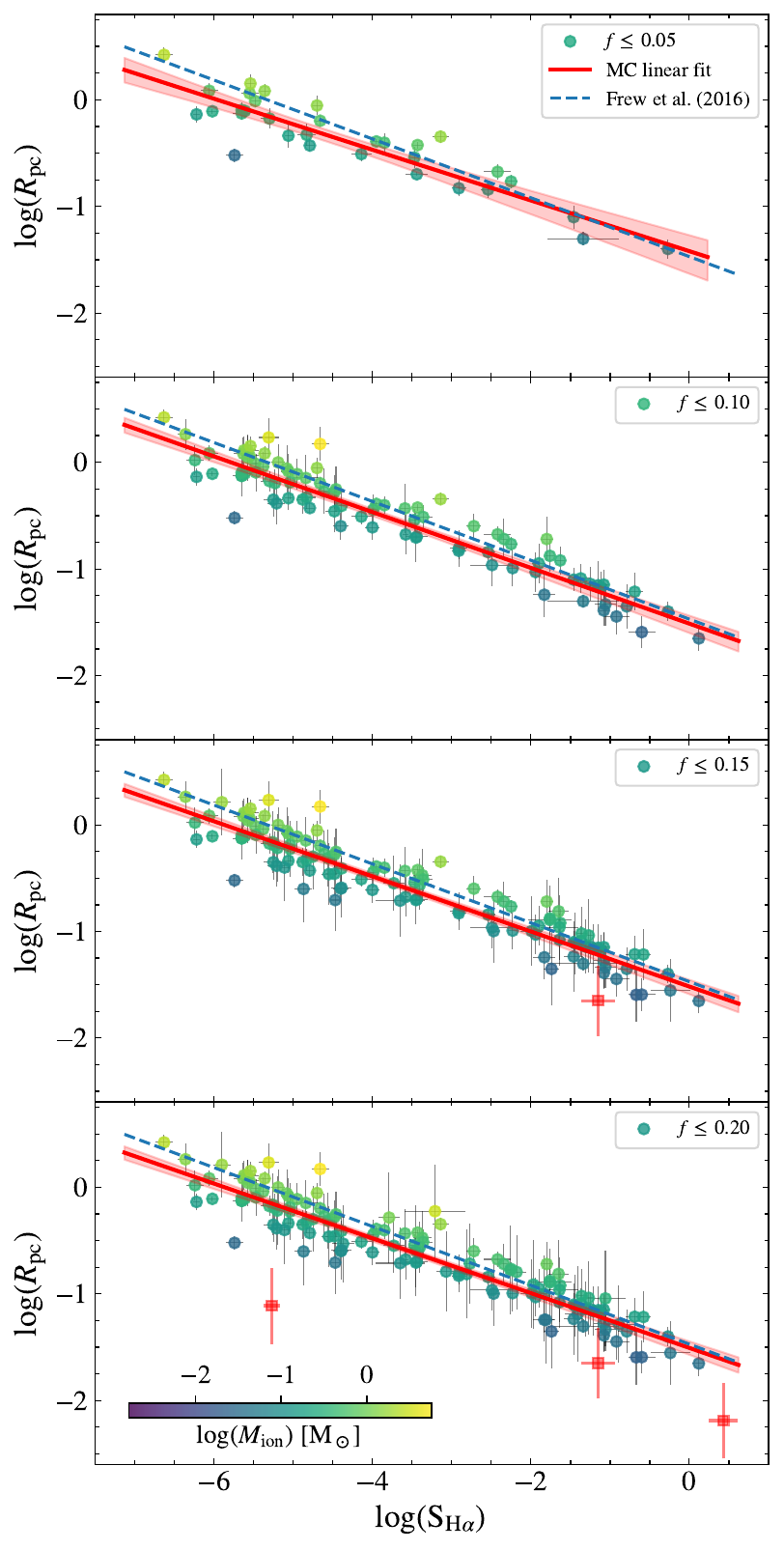} \caption{Recalibration of the 
H$\alpha$ surface brightness--radius relation based on \textit{Gaia} DR3 parallaxes for different fractional parallaxes
uncertainties $f$, as labeled. The red continuous line and the orange shaded region represent the linear fit using a
Monte Carlo procedure and the 95\% confidence interval, respectively. The blue dashed lines represent the linear fit
obtained by \citetalias{frew16} for reference. The data are color-coded by the ionized mass and the PNe with
$\log(M_{\rm ion}) < -2$ (red squares) were excluded from the fit. \label{fig:r_sha_frac_erro}} \end{figure}

We tested the effects of different parallaxes precision by selecting the PNe with fractional parallax uncertainties $f =
\parallaxsd/\parallax$ less than or equal to 0.05, 0.10, 0.15 and 0.20. The results are displayed in Fig.
\ref{fig:r_sha_frac_erro}, where the data are color-coded by the ionized mass, $M_{\rm ion}$, calculated from the 
H$\alpha$ fluxes using the equation provided by \citetalias{frew16}. In the same figure we plotted the
\citetalias{frew16} relation obtained from their full sample of 322 PNe. We performed an Ordinary Least Squares
(OLS) regression for each of the samples using a Monte Carlo (MC) procedure to take into account the data
uncertainties (see Section \ref{appendix_a} in the Appendix for more details). For the MC linear fit, the PNe with
$M_{\rm ion}< 0.01\,M_\odot$ were excluded from the analysis, as the the model assumption of fully ionized, constant
mass on the physical radius--surface brightness distance scale does not strictly hold for these cases
\citepalias{bucciarelli23}. The resulting fitted parameters are summarized in Table \ref{tab:scales}, where the first
column corresponds to the fractional parallax uncertainty upper limit, the second column the number of calibrators, the
third and fourth columns the slope and the intercept of the linear fit with the respective errors, and the fifth column
the Pearson correlation coefficient. We also indicate in the table the averaged distance ratio $\left<k\right>$, which
consists in the average of the statistical distance multiplied by the parallax, $\left<k\right> = \left< D_{\rm stat}
\times \parallax \right>$, and its dispersion $\left<\sigma\right>$, which is an indicator of the goodness of the scale
\citep[see][]{smith15}. As presented in Fig. \ref{fig:r_sha_frac_erro}, independently of the adopted fractional parallax
uncertainty, the linear fits obtained using \textit{Gaia} data are slightly flatter than the obtained from
\citetalias{frew16}, considering the 95\% confidence interval. This results in shorter final statistical distances when
our calibrated relation is used.

\begin{figure}[ht!] \centering \includegraphics[width = 1.0\columnwidth]{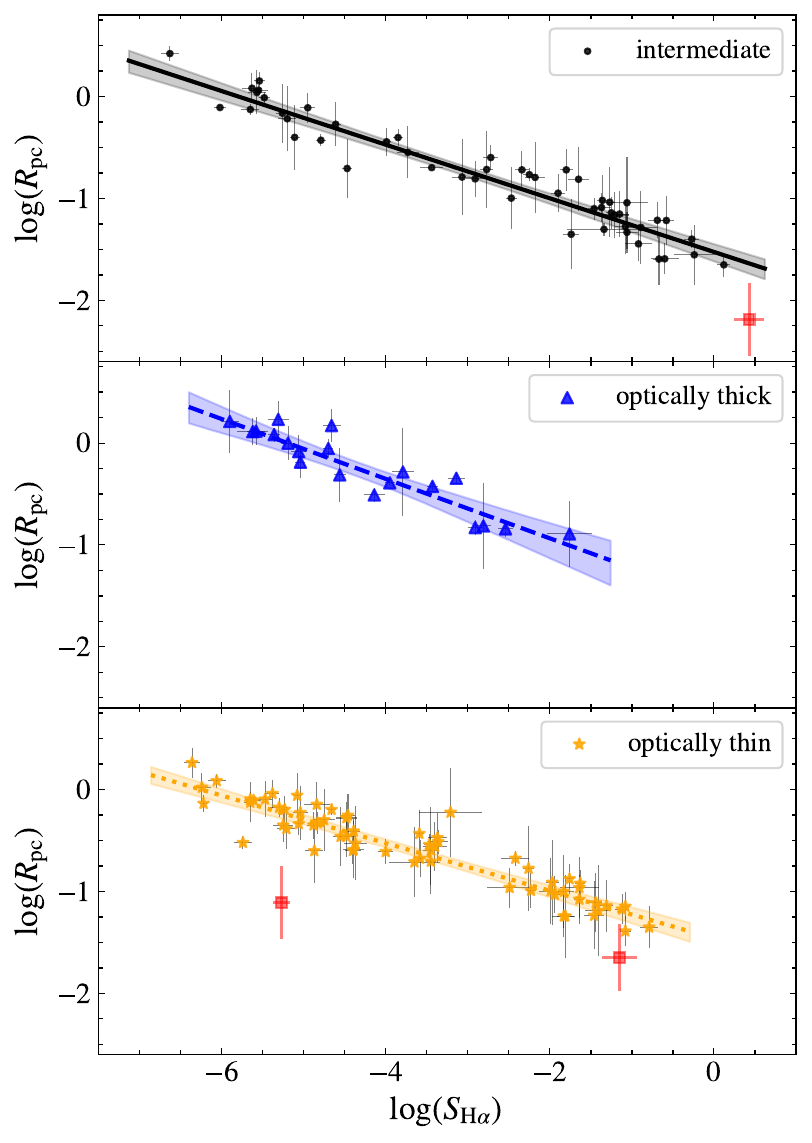} \caption{ The  H$\alpha$ surface
brightness--radius relation based on \textit{Gaia} DR3 parallaxes and classifying the PNe according to the optical
depth: optically thin, thick and intermediate cases, as indicated in each panel. The lines represent linear fits
obtained using a Monte Carlo procedure, as in Fig. \ref{fig:r_sha_frac_erro} and the shaded regions represent the 95\%
confidence intervals for each sub-class. The data are restricted to parallaxes uncertainties $f \le 0.20$ and the PNe
with $\log(M_{\rm ion}) < -2$ (red squares) were excluded from the fit. \label{fig:r_sha_optically}} \end{figure}

As given in Table \ref{tab:scales}, the fitted parameters presented are compatible with each other, given the errors.
The flatter relation is obtained for $f \le 0.05$, with only 32 calibrators and with a lower correlation coefficient
than the other cases. The cases with parallax uncertainties of 0.10, 0.15 and 0.20 provide better fits, with slightly
larger correlation coefficients, compatible with those obtained by \citetalias{frew16}. However, $f \le  0.20$ provides
similar $\left<\sigma\right>$  and $\left<k\right>$ but with 136 calibrators. Therefore, we will consider the case  $f
\le 0.20$  as the best fitted scale and the adopted  H$\alpha$ surface brightness--radius relation as: \begin{equation}
\log R_{\rm pc} = (-0.257 \pm 0.008) \times \log S_{{\rm H}\alpha} -  (1.51 \pm 0.03). \label{eq:halpha_radius}
\end{equation}

A intrinsic dispersion of 0.116 dex about the relation is estimated using a maximum likelihood method and is in
agreement with previous values reported by \citetalias{frew16,bucciarelli23}. Compared to other distance scales
available in the literature, the new calibration based on the $S_{{\rm H}\alpha}$ and \textit{Gaia} DR3 data has been
improved, with $\left<k\right> = 1.020$  and $\left<\sigma\right> = 0.142$ for the $f \le 0.20$ adopted scale. For a
comparison, we have recalculated the averaged distance ratio and its dispersion considering the \citetalias{frew16}
calibration and considering the same data as in the $f \le 0.20$ case from our work, obtaining $\left<k\right> = 1.293$ 
and $\left<\sigma\right> = 0.221$. The values reported by \citetalias{bucciarelli23}, based on the H$\beta$ surface
brightness and the \textit{Gaia} DR3 data, are $\left<k\right> = 0.964$  and $\left<\sigma\right> = 0.154$ for their $f
\le 0.20$ case and with 133 calibrators. This comparison demonstrates the improvement in the current calibration of the
distance scale based on the H$\alpha$ surface brightness--radius relation and \textit{Gaia} DR3 data.

\citetalias{frew16} also notice a dependence of the ${\log(S_{{\rm H}\alpha}})$ -- $\log(R_{\rm pc})$ relation with the
optical depth, resulting in optically thick PNe tending to populate the upper bound of the trend, while optically thin
PNe falling along the lower boundary in the $S_{{\rm H}\alpha}$ -- $R_{\rm pc}$ plane. We follow \citetalias{frew16} and
use the same classification as they use for optically thick and optically thin PNe, based on the intensity of the
spectroscopic lines. Fig. \ref{fig:r_sha_optically} shows the H$\alpha$ surface brightness--radius relation but for PNe
separated in optically thin, optically thick and intermediate cases and considering the fractional parallax uncertainty
$f \le 0.20$.  We also performed MC linear straight line fits and the results are reported in Table \ref{tab:scales}, as
indicated. The differences in the slopes for the three classifications are within the errors, but generally the results
agree with \citetalias{frew16}. Optically thin PNe show a flatter slope and the slope for the optically thick PNe is
steeper. However, as the number of calibrators is not high for each classification, specially for optically thick PNe,
we prefer to use the linear relation for all PNe as presented in equation \ref{eq:halpha_radius} and let the division by
optical depth based on \textit{Gaia} data for future studies with larger datasets.

\begin{deluxetable}{ccccccc} \label{tab:scales} \tabletypesize{\scriptsize} \tablewidth{0pt} \tablecaption{Parameters of
the scales.} \tablehead{\colhead{$f$} & \colhead{$N_\mathrm{cal}$} & \colhead{slope} & \colhead{intercept} &
\colhead{$r$} & \colhead{$\left<k\right>$} & \colhead{$\left<\sigma\right>$}} \colnumbers \startdata
\multicolumn{7}{c}{All}\\ 0.05 & 32 & $-0.239\pm 0.019$ & $-1.43\pm0.09$ & $-0.92$ & 1.025 & 0.246 \\ 0.10 & 88 &
$-0.261 \pm 0.010$ & $-1.51\pm0.04$ & $-0.95$ & 1.017 & 0.141 \\ 0.15 & 120 & $-0.259 \pm 0.008$ & $-1.52\pm 0.03$ &
$-0.95$ & 1.019 & 0.141\\ 0.20 & 135 & $-0.257 \pm 0.008$ & $-1.51\pm0.03$ & $-0.94$ & 1.020 & 0.142\\ \tableline
\multicolumn{7}{c}{intermediate}\\ 0.20 & 51 & $-0.263 \pm 0.012$ & $-1.53\pm 0.04$ & $-0.96$ & 1.012 & 0.154\\
\tableline \multicolumn{7}{c}{optically thin}\\ 0.20 & 64 & $-0.234 \pm 0.012$ & $-1.46\pm 0.05$ & $-0.94$ & 1.007 &
0.151\\ \tableline \multicolumn{7}{c}{optically thick}\\ 0.20 & 20 & $-0.293 \pm 0.035$ & $-1.53\pm 0.16$ & $-0.93$ &
1.001 & 0.388\\
\enddata \end{deluxetable}

\subsubsection{Bayesian distances \label{sec:bayes}}

While the recalibration of the  $\log R_{\rm pc}$\,--\,$\log S_{\rm H\alpha}$ relation is important to obtain
statistical distances for a large set of PNe in the Galaxy, more reliable distances can be obtained for PNe using a
probabilistic approach. \citet{bailer-jones21} estimated stellar distances using a prior constructed from a
three-dimensional model of our Galaxy. Their model includes interstellar extinction and \textit{Gaia} variable magnitude
limit and inferred two types of distances. The geometric uses the parallax with a direction-dependent prior on distance.
The photogeometric, additionally uses the color and apparent magnitude of a star, by exploiting the fact that stars of a
given color have a restricted range of probable absolute magnitudes.  The geometric distances are very dependent on the
Galaxy model adopted, while photogeometric distances depend upon a model of the direction-dependent distribution of
extincted stellar absolute magnitudes. The application of this method to PNe is very uncertain, since CSPNe have a wide
range of properties, with temperatures ranging from $\sim 25,000$ to over 200,000 K, luminosities from 10 to over 10,000
$L_\sun$, and a great variety of spectra \citep{weidmann20}.

The best approach for PNe is to derive distances based on a Bayesian statistics, using the statistical distances and its
uncertainty to define a log-normal prior $P^*(r)$. The \textit{Gaia} parallax \parallax\, and its uncertainty
\parallaxsd\, provide the likelihood. \citetalias{chornay21} used the statistical distances from \citetalias{frew16} as
a prior to derive Bayesian distances and using \textit{Gaia} EDR3 parallaxes. They demonstrated that the relative
uncertainty improvement for the posterior over the statistical distance prior is 1.4, with one-third of the sample
having its relative uncertainties improved by a factor of two or more.  We will proceed as \citetalias{chornay21}, but
instead we will use our recalibration of the $\log R_{\rm pc}$\,--\,$\log S_{\rm H\alpha}$  relation based on
\textit{Gaia} DR3 data to derive more reliable distances for the PNe in our sample. In Section \ref{appendix_b} in the
Appendix we implemented an MC simulation to compare the accuracy of the distances derived with the Bayesian method and
distances derived using only the $S_{{\rm H}\alpha}$ statistical distances, demonstrating that the former have lower
scatter and bias around the true simulated distances for $f < 0.5$.

To estimate the uncertainty, \citet{bailer-jones15} recommends using 5\% and 95\% quantiles to define the 90\%
confidence interval of the posterior distribution. However, to be consistent with previous works that derived distances
for PNe, we report the median and the 16th and 84th percentiles. Because many of our posteriors are not well
approximated as Gaussian, we compute the posterior distance distribution on a fixed grid with 0.01 pc steps between 0
and 50 kpc, and the distributions are normalized by the sum of values over the domain of the prior. For PNe lacking
\textit{Gaia} DR3 parallaxes, the posterior distributions were adopted as the prior distributions. More details can be
seen in Section \ref{appendix_b} in the Appendix.

\begin{figure} \centering \includegraphics[width = 1.0\columnwidth]{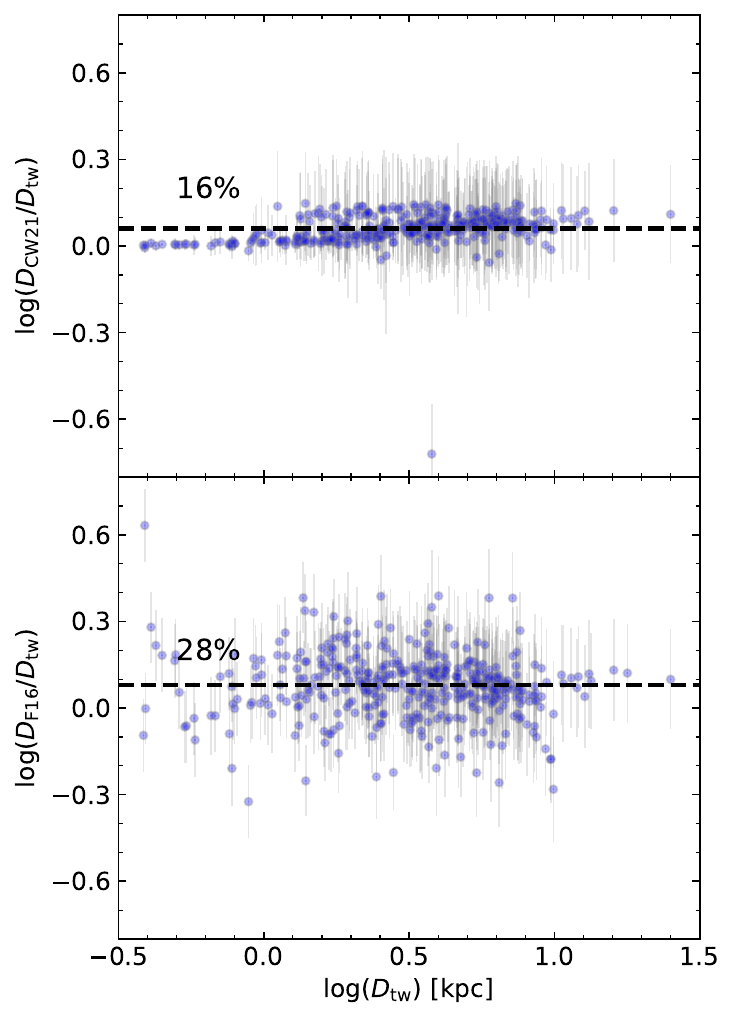} \caption{Top panel: Comparison between
heliocentric distances derived from this work ($D_{\rm tw}$) using the Bayesian approach for PNe with available
\textit{Gaia} DR3 parallaxes and the ratio of the heliocentric distances from \citet{chornay21} and this work, in
logarithm for a better visualization. Bottom panel: the same as in the top panel but for \citetalias{frew16} distances.
In both panels the horizontal dashed lines indicate the average difference in the compared distances: 16\% for
\citetalias{chornay21} and 28\% for \citetalias{frew16}. \label{fig:comp_dist_chornay}} \end{figure}

Comparisons of the obtained Bayesian distances for PNe with available \textit{Gaia} DR3 parallaxes following this method
are presented in Fig. \ref{fig:comp_dist_chornay}, where in the top panel a comparison with \citetalias{chornay21} is
provided. \citetalias{chornay21} distances are on average 16\% larger than our Bayesian distances. We also note the
effect of the recalibration of the statistical scale performed in this work, as for PNe closer to the Sun (distances
smaller than $\sim$1500 pc) the spread in the data considering $\log(D_{\rm CW21}/D_{tw})$ is much lower.
\citetalias{frew16} distances are on average 28\% larger, as can be seen in the bottom panel of the figure. For this
case, the Bayesian method also increases the scatter in the distances ratio, as the distances also rely on the
trigonometric parallaxes. Therefore, the distances provided in this work are shorter than previous distances based on
the H$\alpha$ surface brightness--radius relation.

We recalculated the mean distance ratio and its dispersion for the calibration sample used in Fig.
\ref{fig:r_sha_frac_erro} considering the $f \le 0.20$ case and using the Bayesian distances, obtaining $\left<k\right>
= 1.007$  and $\left<\sigma\right> = 0.019$. This result demonstrates the validity of the Bayesian distances based on
the H$\alpha$ surface brightness--radius relation and \textit{Gaia} DR3 data.

The final distances adopted for the 1,130 PNe are Bayesian distances based on the $S_{\textrm{H}\alpha}$ statistical
scale. For a subsample of 415 PNe with measured \textit{Gaia} parallaxes the Bayesian distances use a prior based on the
$S_{\textrm{H}\alpha}$ statistical distances and the likelihood  computed from the parallaxes. For the remaining 711 PNe
without \textit{Gaia} counterparts, the posterior distribution is adopted as the prior distribution. The heliocentric
distances derived in this work are converted to Galactocentric distances $R$ using the \texttt{Astropy} package. For
this, the distribution of the heliocentric distance obtained from the posterior is converted for Galactocentric
distances and, from the resulted distribution, we compute the median and the 16th and 84th percentiles, obtaining the
Galactocentric distance and its uncertainty.

While the  $S_{\textrm{H}\alpha}$ statistical distance scale is extensively compared with other distances scales by
\citetalias{frew16}, in Section \ref{appendix_c} in the Appendix we compare our recalibrated  $S_{\textrm{H}\alpha}$
statistical distances with the most recent statistical distances based on $S_{\textrm{H}\alpha}$ or, equivalently,
$S_{\textrm{H}\beta}$, and also extinction distances. In particular, the effect of the Galactic extinction in
the distances determined from $\textrm{H}\alpha$ and $\textrm{H}\beta$ fluxes is investigated in Section
\ref{appendix_c1}. Since these emission lines are produced by the same ion within the same region of the nebula,
both correlate with the nebular physical size. However, according to Case B recombination theory \citep{osterbrock06},
the H$\alpha$ line is approximately three times brighter than the H$\beta$ line, providing higher signal-to-noise ratio
fluxes for deriving statistical distances. Furthermore, H$\beta$ fluxes are more severely affected by interstellar
extinction than H$\alpha$ fluxes, particularly in high-extinction PNe. Therefore, statistical distances based on
H$\alpha$ fluxes are preferred.

\subsection{Velocities and halo, thin and thick disks selection criterion}

Some of the \textit{Gaia} DR3 sources have the equatorial proper motions and errors $\mu_{\alpha^*} \pm
\sigma_{\mu_{\alpha^*}}$ and  $\mu_\delta \pm \sigma_{\mu_\delta}$ provided by the satellite. Combining these proper
motions with the radial velocities from \citet{durand98}, we are able to obtain the 3D spatial motions of some PNe in
our sample. We proceed as in \citetalias{bucciarelli23} to convert the DR3 proper motions and their standard deviations
into Galactic coordinates and to compute the corresponding observed spatial velocities plus errors: $V_\ell \pm
\sigma_{V_\ell}$ and $V_b \pm \sigma_{V_b}$, where $\ell$ and $b$ are the Galactic longitude and latitude. The peculiar
velocity ($V_{\rm pec}$) is the residual stellar motion that deviates from the general Galactic rotation. The procedures
to obtain $V_{\rm pec}$, the radial ($V_R$), azimuthal ($V_\phi$) and vertical ($V_z$) components of the spatial
velocity are calculated as in \citetalias{bucciarelli23} and we refer the reader to their section 3.2 for more details.
In this work we adopt the Sun distance to the Galactic Center $R_\sun = 8.122$~kpc from \citet{gravity2018}.

\begin{figure}[ht!] \includegraphics[width = 1.0\columnwidth]{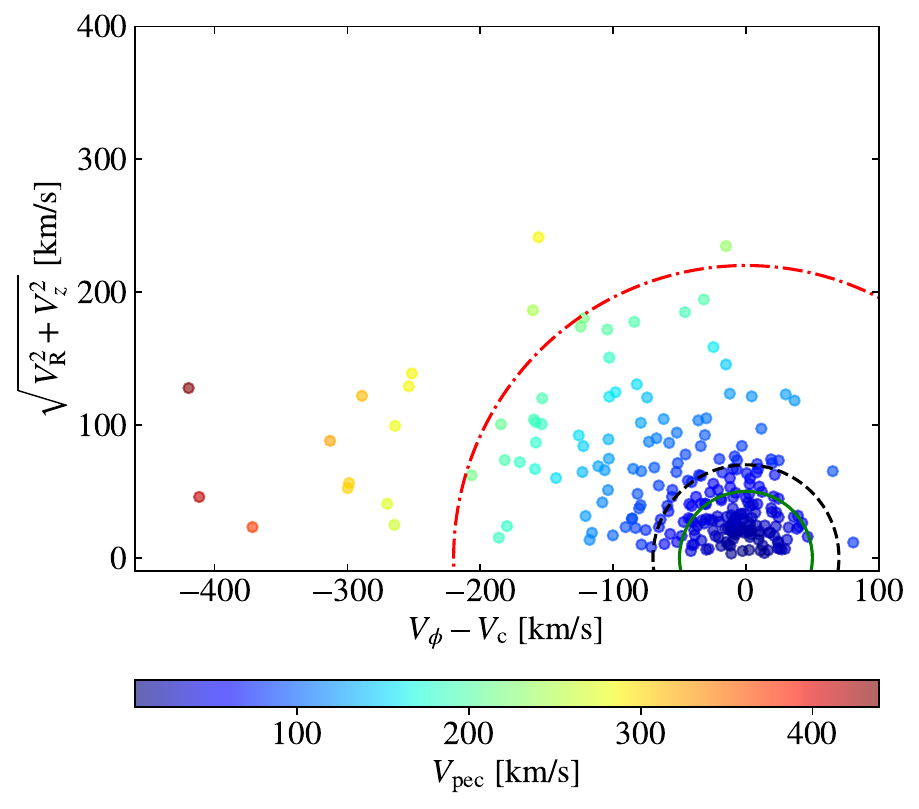} \caption{Toomre diagram to identify halo, thin and
thick disks PNe. Green, black and red semi-circles show constant values of the total Galactic velocities of 50, 70, and
220 km/s. The data are color-coded using the $V_{\rm pec}$ velocities. \label{fig:toomre_diag}} \end{figure}

As the matching methodology for \textit{Gaia} sources and CSPNe is not the same and also the distances derived in this
work are derived from a recalibration of the $\log R_{\rm pc}$\,--\,$\log S_{\rm H\alpha}$ relation and using the
Bayesian approach, the velocities derived by us are not exactly the same as those from \citetalias{bucciarelli23}. In
order to select the PNe belonging to the halo and the thin and thick disks, or transitional phases, we constructed the
Toomre diagram, as shown in Fig. \ref{fig:toomre_diag}, as in \citet{bensby14}. Thin disk PNe population have $V_{\rm
pec}$ lower than 50 km/s and a thick disk stars with velocities between 70 and 220 km/s. Stars with velocities between
50 and 70 km/s are the ones considered to be transitory members between the thin and thick disks. Halo PNe have $V_{\rm
pec}$ higher than 220 km/s.

\subsection{Oxygen abundances}

The oxygen abundances used in this work are from \citet{stanghellini18} and \citetalias{bucciarelli23}. They compiled a
large set of chemical abundances from the literature and recalculate them using the same set of ionization correction
factor (ICF) from \citet{kingsburgh94}. While the recalculation of the chemical abundances with the same set of ICF is
essential, we still expect some dispersion of the data due to differences in the atomic data, specific lines and
density and temperature diagnostics utilized by the different authors. For O$^{++}$ the expected average abundance
differences are of the order of 0.01 dex, with some larger divergences of 0.05 dex \citep[see][]{stanghellini18}.
\citet{henry10} utilized a large database of 124 PNe with homogeneously determined abundances to study the
oxygen abundance gradient in the Galactic disk. The chemical abundances in their work were determined with the same
atomic data, plasma diagnostic scheme and ionic abundance calculations. Oxygen elemental abundances are calculated using
the ICF from \citet{kingsburgh94}. To test for possible large variations of chemical abundances within the data utilized
in this work, we compared the chemical abundances for 71 PNe of \citet{stanghellini18} and \citetalias{bucciarelli23}
with 42 PNe from \citet{henry10} by selecting a subset of PNe in the solar neighborhood with galactocentric distances in
the range $7.0 \le R \le 9.0$\,kpc in both samples. Then, we computed the mean and the standard deviation of the
$\log(\mbox{O}/\mbox{H}) + 12$ abundances, obtaining 8.61 dex and 0.17 dex for \citet{henry10} data and 8.63 dex and
0.21 dex for \citet{stanghellini18} and \citetalias{bucciarelli23} data, respectively. The mean and the standard
deviation for the \citet{stanghellini18} and \citetalias{bucciarelli23} sample are slightly larger than the ones for
\citet{henry10}, however the difference of the standard deviations is 0.04 dex, smaller than the expected uncertainties.
Therefore, we can rule out the possibility of large variations between the different sources of chemical abundances used
in this work.

As PNe are the product of the evolution of 1--8 M$_\sun$ stars, the $\alpha$-elements abundances are not expected to be
modified during the evolution of the progenitor stars, reflecting the conditions of the ISM at the time the progenitors
were formed. However, for oxygen a small production or depletion may be observed due to both the dredge-up (DU)
episodes and the hot-bottom burning (HBB) during the AGB phase \citep{ventura15}, that depend on the
metallicity and the mass of progenitor star. The modifications may be more relevant for the most massive progenitor
stars, which are rare in the PNe population \citep[see a discussion by][]{stanghellini24}.
\citet{delgado-inglada15} find evidence for oxygen enrichment by $\sim$0.3 dex for a small sample of seven
carbon-rich (circumstellar) dust (CRDs) PNe of intermediate metallicities in the range $\log(\mbox{O}/\mbox{H}) + 12 = $
8.2 -- 8.7, while oxygen is invariant in PNe with oxygen-rich (circumstellar) dust (ORDs). They suggest that chlorine
(preferentially) and argon should be used as metallicity probes instead of oxygen. The theoretical interpretation for
this difference is the extra-mixing from the convective shell \citep{garcia-hernandez16}. On the other hand, many
studies point to oxygen varying in lockstep with other $\alpha$-elements in PNe, specially argon, as e.g.
\citet{milingo10,maciel17, cavichia17,kwitter22,tan24}.  \citet{maas21} compared chlorine and oxygen abundances from
\citet{delgado-inglada15} and \citet{henry04}, selecting PNe with galactocentric radii between 6 and 11 kpc. They find
good agreement when comparing with abundances from 52 M giants, which would indicate that both chlorine and oxygen in
PNe are good ISM metallicity probes. \citet{bhattacharya22} analyze the distribution of oxygen abundances against the
argon abundances of 101 Milky Way PNe, whose dust properties and abundances were tabulated by \citet{ventura17}. They
conclude that there is no segregation of CRD/ORD PNe. However they note that mixed chemistry dust (MCDs) PNe that are
metal-rich may indicate oxygen depletion. These PNe likely evolve from the most massive progenitors that display HBB
\citep{ventura17} and represent a small fraction of Galactic PNe \citep{stanghellini24}.

Consequently, the expected variations in the oxygen abundances of PNe are small compared to current measurement
uncertainties. However, there are exceptions such as CRDs and MCDs PNe, which may not serve as reliable tracers of ISM
metallicity. Throughout this work we adopt oxygen as a proxy for the ISM metallicity, subject to the previously
discussed caveats. To derive the radial abundance gradients, we consider the oxygen abundance ratio O/H, which has been
derived for a large number of objects in the Galaxy and has the smallest uncertainties among all known abundances.

\section{Results \label{sec:results}}

The results reported in the previous section are summarized in Table \ref{tab:dist_catalog}, where we provide the
heliocentric distances calculated with the recalibrated $\log R_{\rm pc}$\,--\,$\log S_{\rm H\alpha}$ relation and using
the Bayesian method. Peculiar velocities $V_{\rm pec}$, Galactocentric distances as their components are also reported
in the table. Asterisks in the ``\textit{Gaia} id.'' column mark PNe with central stars identified in the \textit{Gaia}
DR3 data archive. Column descriptions are provided in the table notes.

\begin{deluxetable*}{llccccccccccc}
\tablewidth{0pt}
\setlength{\tabcolsep}{1.7pt}
\tablecaption{Catalog of distances.\label{tab:dist_catalog}}
\tablehead{
\colhead{PNG} & \colhead{Name}  & \colhead{$\theta$}  & \colhead{$\log(S_{\mathrm{H}\alpha})$} & \colhead{E(B-V)} & \colhead{$V_{\mathrm{pec}}$} & \colhead{$D$} & \colhead{$R$} & \colhead{$R_x$} & \colhead{$R_y$} & \colhead{$R_z$} & \colhead{Flag} & \colhead{O/H} \\
\colhead{}   & \colhead{}  & \colhead{(arcsec)} & \colhead{($\mbox{cgs}\,\mbox{sr}^{-1}$)}  & \colhead{(mag)}  & \colhead{(km/s)} & \colhead{(kpc)} & \colhead{(kpc)}  & \colhead{(kpc)}  & \colhead{(kpc)}  & \colhead{(kpc)}  & \colhead{}  & \colhead{(dex)} \\
\colhead{(a)} & \colhead{(b)} & \colhead{(c)} & \colhead{(d)} & \colhead{(e)} & \colhead{(f)} & \colhead{(g)} & \colhead{(h)} & \colhead{(i)} & \colhead{(j)} & \colhead{(k)} & \colhead{(l)} & \colhead{(m)}
}
\startdata
000.0-06.8 & H 1-62 & 2.24 & $-1.25\pm0.29$ & $0.49\pm0.29$ & $190.6\pm42.3$ & $8.47^{+1.97}_{-1.45}$ & $1.11^{+1.33}_{-0.77}$ & $0.30^{+2.00}_{-1.47}$ & $0.00^{+0.00}_{-0.00}$ & $-1.01^{+0.18}_{-0.25}$ & * & \nodata \\
000.1+17.2 & PC 12 & 1.12 & $-0.65\pm0.32$ & $0.54\pm0.31$ & \nodata & $8.32^{+3.25}_{-2.34}$ & $1.76^{+1.87}_{-1.21}$ & $-0.16^{+3.07}_{-2.21}$ & $0.02^{+0.01}_{-0.01}$ & $2.47^{+0.94}_{-0.68}$ & \nodata & \nodata \\
000.1-01.7 & PHR J1752-2941 & 7.14 & $-3.07\pm0.33$ & $0.99\pm0.31$ & \nodata & $5.49^{+2.17}_{-1.55}$ & $2.72^{+1.50}_{-1.70}$ & $-2.64^{+2.17}_{-1.54}$ & $0.01^{+0.01}_{-0.00}$ & $-0.16^{+0.05}_{-0.07}$ & \nodata & \nodata \\
000.1-02.3 & Bl 3-10 & 3.52 & $-2.41\pm0.30$ & $0.64\pm0.25$ & \nodata & $7.53^{+2.87}_{-2.08}$ & $1.78^{+1.73}_{-1.22}$ & $-0.59^{+2.91}_{-2.10}$ & $0.03^{+0.01}_{-0.01}$ & $-0.31^{+0.09}_{-0.13}$ & \nodata & \nodata \\
000.1-05.6 & H 2-40 & 8.79 & $-3.22\pm0.23$ & $0.50\pm0.22$ & \nodata & $4.87^{+1.72}_{-1.27}$ & $3.27^{+1.25}_{-1.63}$ & $-3.27^{+1.69}_{-1.25}$ & $0.01^{+0.00}_{-0.00}$ & $-0.47^{+0.13}_{-0.17}$ & \nodata & \nodata \\
\enddata
\tablecomments{Table \ref{tab:dist_catalog} is published in its entirety in the machine-readable format and available electronically at the CDS. A portion is shown here for guidance regarding its form and content. The columns of the table correspond to the following identifications:  (a) PN PNG number; (b) PN name; (c) angular radius in arcsec; (d) H$\alpha$ surface brightness in $\mbox{erg}\,\mbox{cm}^{-2}\mbox{s}^{-1}\mbox{sr}^{-1}$; (e) reddening in magnitudes; (f) peculiar velocity in km/s; (g) heliocentric distance in kpc; (h) Galactocentric distance in kpc; (i) $x$-component of the Galactocentric distance in kpc; (j) $y$-component of the Galactocentric distance in kpc; (k) $z$-component of the Galactocentric distance in kpc; (l) \textit{Gaia} CSPN identification flag; (m) oxygen abundances $\log({\rm O}/{\rm H})+12$ in dex.}
\end{deluxetable*}

\subsection{2D distribution of the PNe in the Galactic plane}

\begin{figure*} \centering \includegraphics[width = 0.9\textwidth]{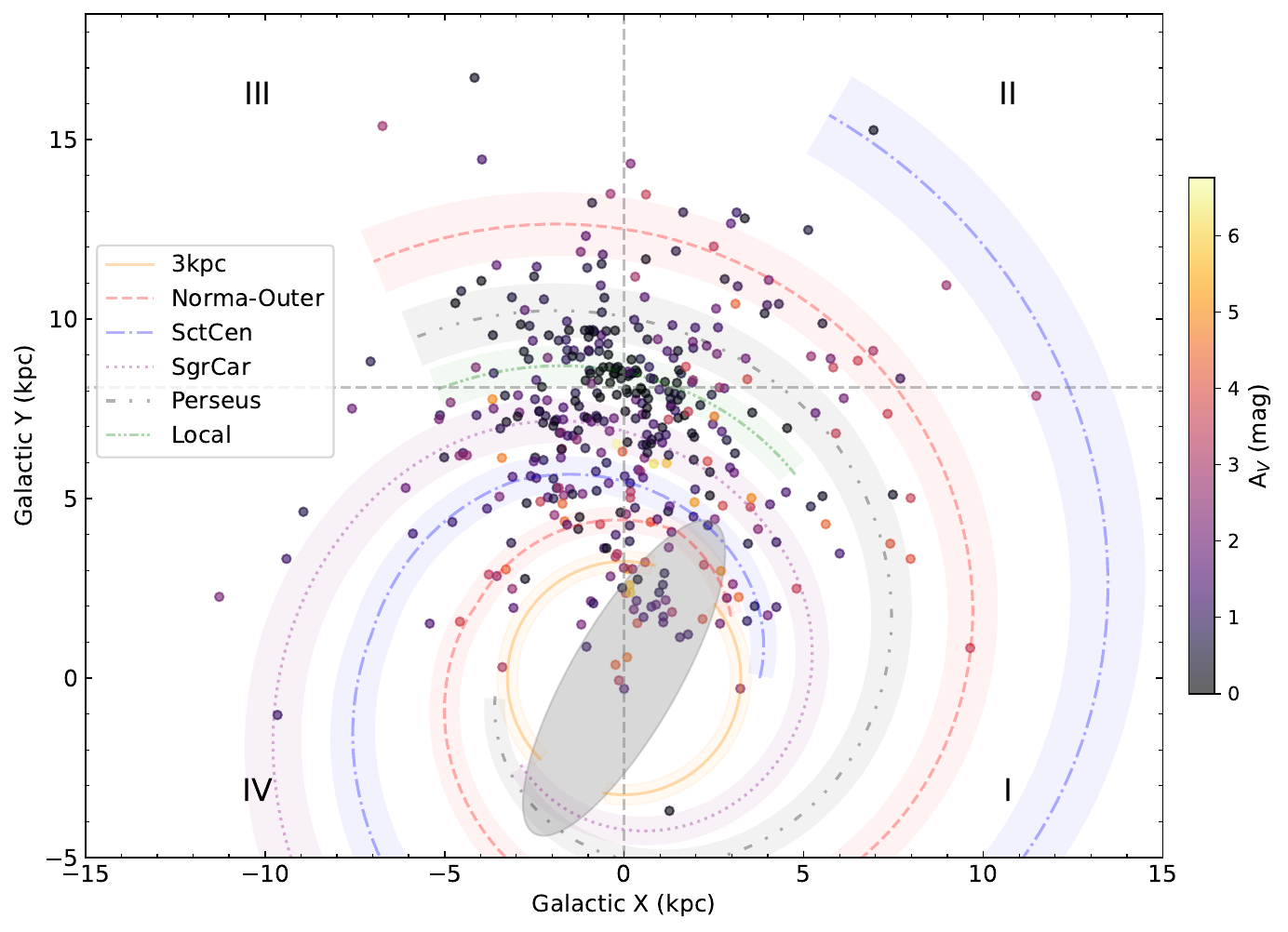} \caption{Distribution of the 415 PNe in the
Galactic plane as seen from top view and using our Bayesian distances calculated in Section \ref{sec:bayes}. The color
bar indicates the interstellar extinction ($A_V$) as described in Section \ref{sec:data}. Milky Way spiral arm
positions are based on \citet{reid19}, with names provided in the legend. The shaded ellipse near the center represent
the position of the Galactic bar after \citet{wegg15}. The location of the Sun is marked by the interception of the
dashed lines at (0.0,8.1) and the Galactic Center is located at (0,0). The Galactic quadrants are marked using roman
numerals. \label{fig:mw_spiral_arms}} \end{figure*}

The 415 PNe with Bayesian distances based on \textit{Gaia} DR3 parallaxes enable a study of the 2D distribution of PNe
in the Galactic plane. Fig. \ref{fig:mw_spiral_arms} shows this distribution, with the spiral arms and Galactic bar
over-plotted for context (see caption). Since PN progenitor stars span a wide range of ages and most of them likely have
ages $<4$\,Gyr \citep{maciel13}, they are not expected to strictly trace the spiral arms. However, the distribution
shown in Fig. 6 reveals concentrations in regions that coincide with the expected locations of the spiral arms. 
Similarly, in the central regions, they are concentrated where the bulge and bar are expected to exist.

In quadrant IV there are PNe located in the Norma-Outer and Scutum-Centaurus arms. We also observe an
overdensity of PNe in the Local arm, near the corotation radius \citep{dias19}. In quadrants I and II the number of PNe
drops beyond the Norma-Outer arm. Beyond the Galactic Center, in the I and IV quadrants, we cannot obtain distances for
the PNe using the Bayesian method, because the parallaxes for these sources cannot be measured. In quadrant III,
three PNe are located far beyond the Norma-Outer arm, which may suggest either radial migration or large distance
uncertainties. The data also capture the the higher interstellar extinction in the direction of the central regions of
the Galaxy, meanwhile in the solar neighborhood the extinctions are lower.

\subsection{Radial oxygen abundance gradients} One of the most important aspects in deriving accurate distances for
Galactic PNe is to study the radial metallicity gradients in the Galaxy. For this purpose, and to obtain the metallicity
gradient for a large Galactocentric distance interval using as many as PNe with accurate chemical abundances available,
in this Section we will use all available distances published in Table \ref{tab:dist_catalog}. This way, most of the
nearest PNe have more accurate distances, based on \textit{Gaia} parallaxes, while the farthest PNe have distances
derived based on the statistical distances.

\begin{figure}[ht!] \includegraphics[width = 1.0\columnwidth]{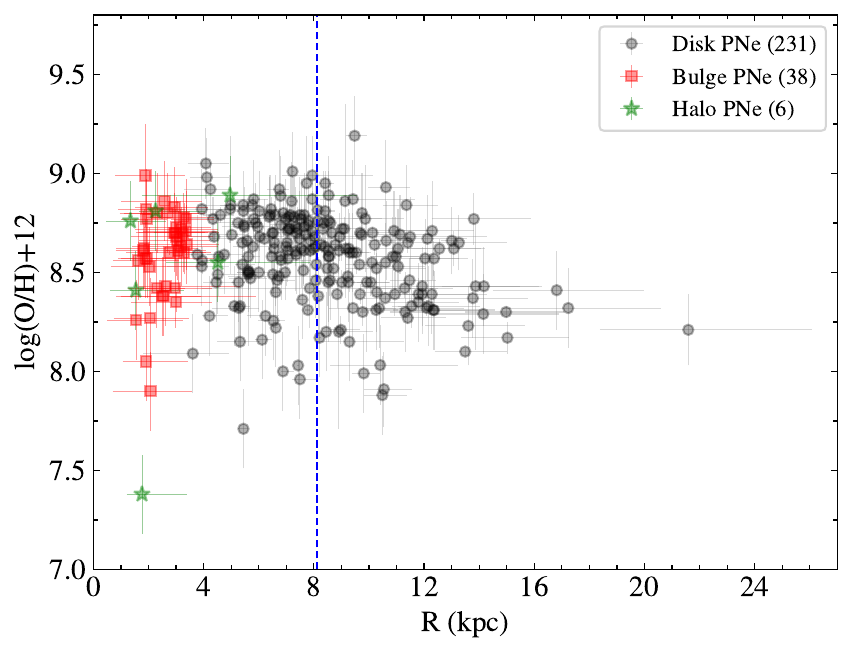} \caption{Radial abundance distribution for oxygen
including all objects in our sample. The PNe were classified as disk, bulge and halo  with the number of PNe in each
population indicated in parenthesis, as labeled. The vertical dashed line marks the Sun position at 8.122 kpc.
\label{fig:disk_grads_scatter_plot}} \end{figure}

The radial O/H distribution is shown in Fig. \ref{fig:disk_grads_scatter_plot}, where the PNe were separated in those
pertaining to the disk (231 PNe), bulge (38 PNe) or halo (6 PNe) populations. As in \citetalias{bucciarelli23}, the halo
population was selected by those PNe with $V_{\rm pec} > 220$\,km/s. The bulge PNe were selected as those PNe with $R <
3.5$\,kpc and are in the direction of the Galactic Center with Galactic coordinates $ 335^{\circ} < \ell < 25^{\circ}$
and $|b| < 25^{\circ}$. The remaining PNe were classified as disk population. Interestingly, inside the solar radius the
data show a flat distribution, while outside the solar radius there is a decrease in the the abundances as the
Galactocentric distances increase, until $\sim 14$\,kpc, where the abundance distribution seems to become flat again.

In order to provide a more objective gradient fitting procedure, we will employ in this work the same methodology as in
\citet{cardoso25}, where a detailed analysis of the O/H radial abundance gradients derived from the H {\sc ii} region is
performed in 154 isolated spiral galaxies observed by the CALIFA survey. The problem of gradient fitting is addressed
using a robust unsupervised automatic fitting procedure that employs a bootstrap process on the data to escape local
minima. To perform the fit of the abundance radial gradient, we use the Python package \texttt{piecewise regression}
\citep{pilgrim21}. This method simultaneously fits breakpoints positions and linear models for the different fit
segments and gives confidence intervals for all the model estimates. The chosen models are a simple linear fit and
piecewise linear functions presenting one or two breaks in the radial distributions.

\begin{figure} \includegraphics[width = 1.0\columnwidth]{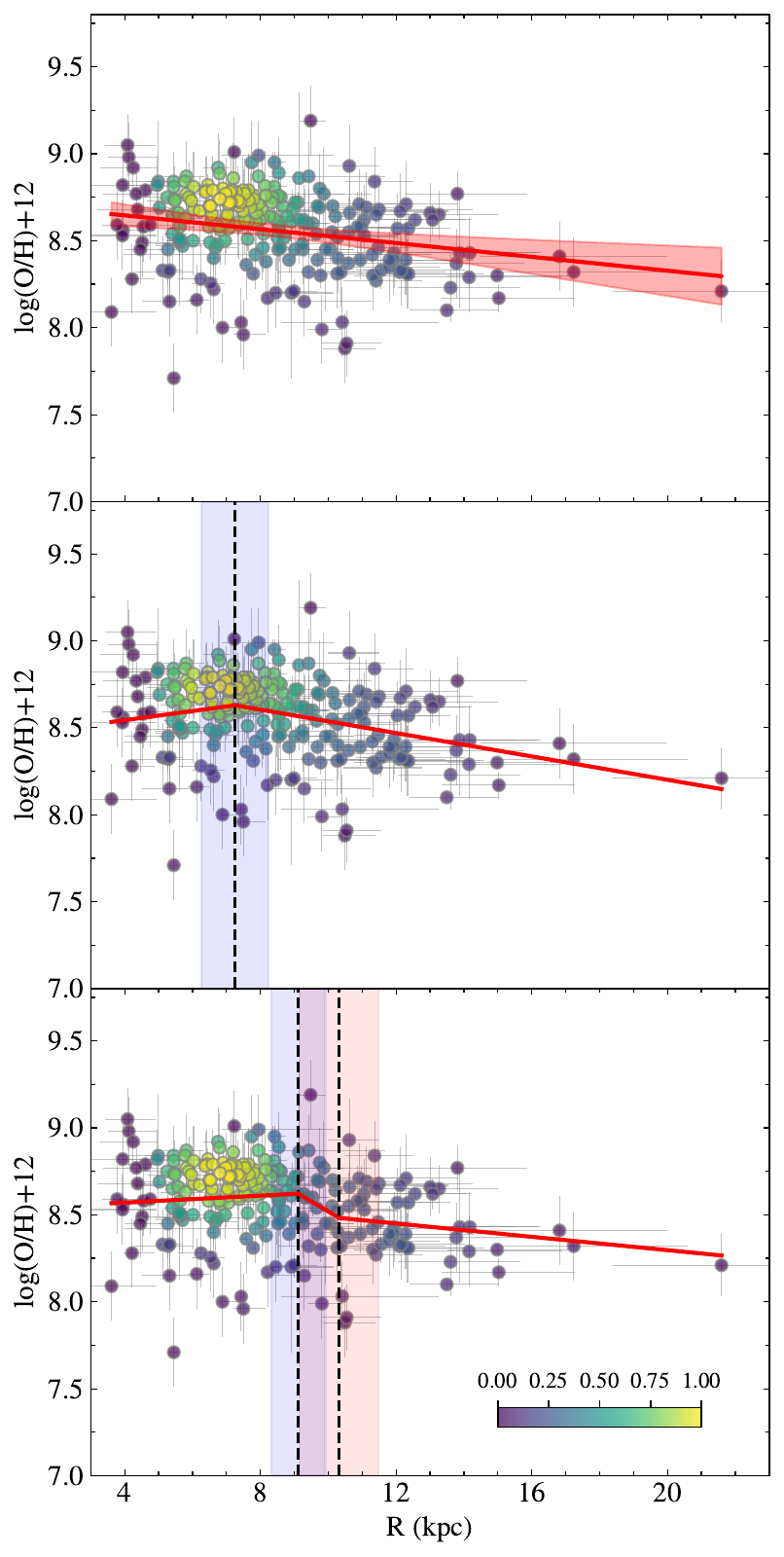} \caption{Radial O/H gradient for disk PNe. Top panel: a
simple linear fit with the shaded region representing the 95\% confidence interval (linear model). Middle panel: a fit
considering a break in the radial distribution (one-break model). Bottom panel: a fit with two breaks (two-break model).
The vertical dashed lines mark the position of the breaks and the shaded regions the 68\% confidence interval.
The color bar indicates the normalized density of points. \label{fig:grads_breaks}} \end{figure}

In the radial abundance gradient analysis we have included only disk PNe, disregarding bulge and halo PNe. Fig.
\ref{fig:grads_breaks} shows the results of the fitted models and Table \ref{tab:grads_breaks} summarizes the
parameters. In this table the columns refer to the model (linear, one-break and two-break) and the lines the fitted
parameters. In the case of the linear model, $a_1$ and $b_1$ represent the slope and the intercept. For the one-break
and two-break models they represent the parameters of the first segment, $a_2$ is the slope of the second segment for
the one-break and two-break models and $a_3$ is the slope of the third segment for the two-break model.

To decide between the models, the Akaike information criterion \citep[AIC,][]{akaike.test} is adopted. We refer to
\citet{cardoso25} and references therein for more information. The second to last row of  Table \ref{tab:grads_breaks}
presents the AIC values for each of the fitted models. The most probable model is the one with one break in the radial
distribution, presenting the lowest AIC value. In this case, there is a break in the radial distribution at $7.25 \pm
1.00$\,kpc, close to the solar radius. Both models with one and two breaks indicate a flatter gradient inside the solar
radius, while outside the solar radius the gradient is steeper. For the one-break model the slopes are $0.026 \pm 0.028$
and $-0.034 \pm 0.009$\,dex/kpc, for the first and second segments, respectively. The two-break model has slopes
$0.010\pm0.015$, $-0.115\pm0.135$ and $-0.019\pm0.018$~dex/kpc, for the first, second and third segments, respectively.
In this case, the second segment is the steepest, however with a very short radial range of about $1$~kpc. Therefore, in
this case there is a step in the radial O/H abundance distribution near the Sun position.

\begin{deluxetable}{cccc} \tabletypesize{\scriptsize} \tablewidth{0pt} \tablecaption{Parameters of the fitted models.
\label{tab:grads_breaks}} \tablehead{ & \multicolumn{3}{c}{Model}\\ \cline{2-4} \colhead{Parameters} & \colhead{Linear}
& \colhead{One-break} & \colhead{Two-break}} \colnumbers \startdata $a_1$ (dex/kpc) & $-0.020 \pm 0.006$ &  $0.026 \pm
0.028$&   $0.010 \pm 0.015$  \\ $a_2$ (dex/kpc) & \nodata & $-0.034 \pm 0.009$ & $-0.115\pm 0.135$\\ $a_3$ (dex/kpc) &
\nodata & \nodata & $-0.019 \pm 0.018$\\ $b_1$ (dex) & $8.73 \pm 0.05$  & $8.44 \pm 0.17$ & $8.53\pm0.10$ \\ $h_1$ (kpc)
& \nodata & $7.25 \pm 1.00$ & $9.11 \pm 0.80$ \\ $h_2$ (kpc) & \nodata & \nodata & $10.32 \pm 1.20$\\ AIC & 35.38 &
31.85 & 33.49\\ $\mathcal{L}(m_i|x)$ & 0.17 & 1 & 0.44 \\ \enddata \end{deluxetable}

The likelihood of a model given the data $\mathcal{L}(m_i|x)$ calculated based on the AIC values
\citep{Burnham2002} are presented in the last row of Table \ref{tab:grads_breaks}, where the minimum AIC model is taken
as reference. For alternative models that $\mathcal{L}(m_i|x) > 0.05$, the selection of the best model is inconclusive.
Given the AIC differences, the linear model fit is notably weaker than the one-break model, being approximately
one-sixth as likely. However, the current PNe data do not allow to choose between the one-break or two-break models. For
this, we need data with smaller uncertainties both in the chemical abundances and distances to confirm the change in
slope of the radial O/H gradient near the solar radius.

\subsection{Azimuthal variations in the metallicity distribution}

To investigate the azimuthal structure in the metallicity distribution, we adopted a procedure similar to \citet[][and
references therein]{wenger19}. They analyzed the metallicity structure in the Galactic disk from \Hii regions and using
the software package \texttt{pyKrige}\footnote{See \url{https://github.com/bsmurphy/PyKrige}.}, which employs kriging
\citep[see][]{feigelson12}, to interpolate the distribution of oxygen abundances and to produce an abundance map in the
Galactic plane. This technique is particularly suitable for PNe, since their distribution across the Galactic disk is
sparse and irregular, and kriging provides an unbiased and variance-minimizing interpolation under such conditions. To
interpolate the data we define a grid of Galactic coordinates in the plane ranging from -12 to 12 kpc in the
$x$-direction and from -4 to 20 kpc in the $y$-direction in steps of 1 kpc. To generate an error map to access the
uncertainties in the procedure, an MC simulation was implemented, which consists of generating random Gaussian values in
the distances and abundances within their uncertainties. We adopted 500 MC  realizations, generating in this way for
each simulation a kriging map using a linear semivariogram model. Then, for each cell of the grid in the Galactic plane,
we construct a probability density function (PDF) of the interpolated values and fit a kernel density
estimation (KDE) to the abundance distributions. The adopted interpolation value of abundance is the peak of the fitted
KDE and the 1-$\sigma$ confidence interval is represented by the bounds of the distribution that encompass 68\% of the
PDF area.

\begin{figure} \includegraphics[width = 1.0\columnwidth]{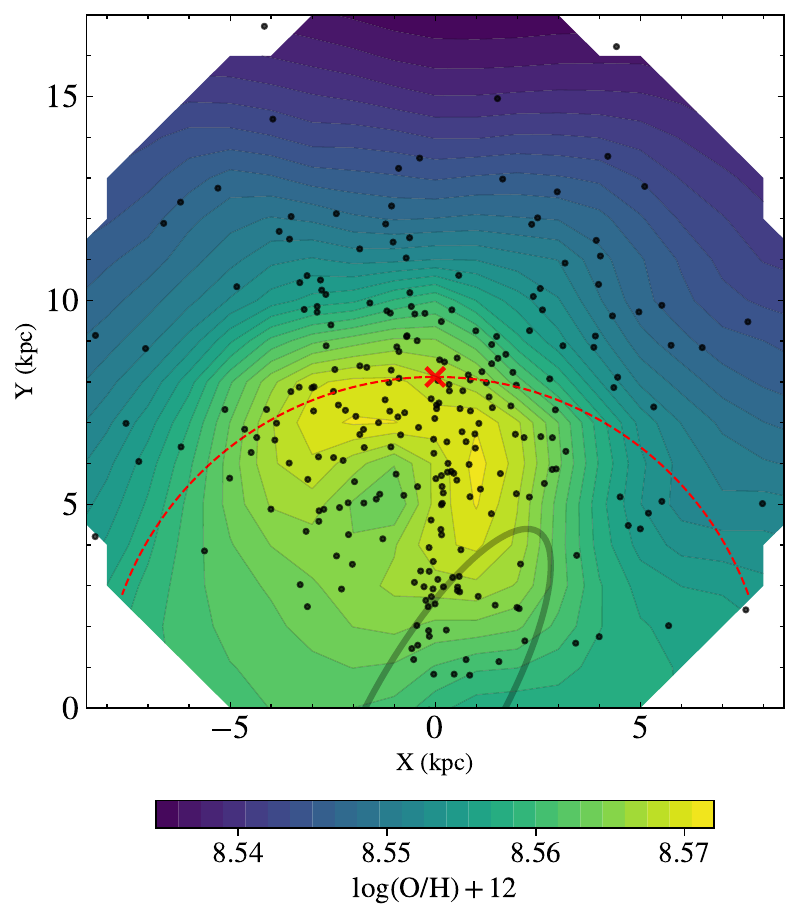} \includegraphics[width = 1.0\columnwidth]{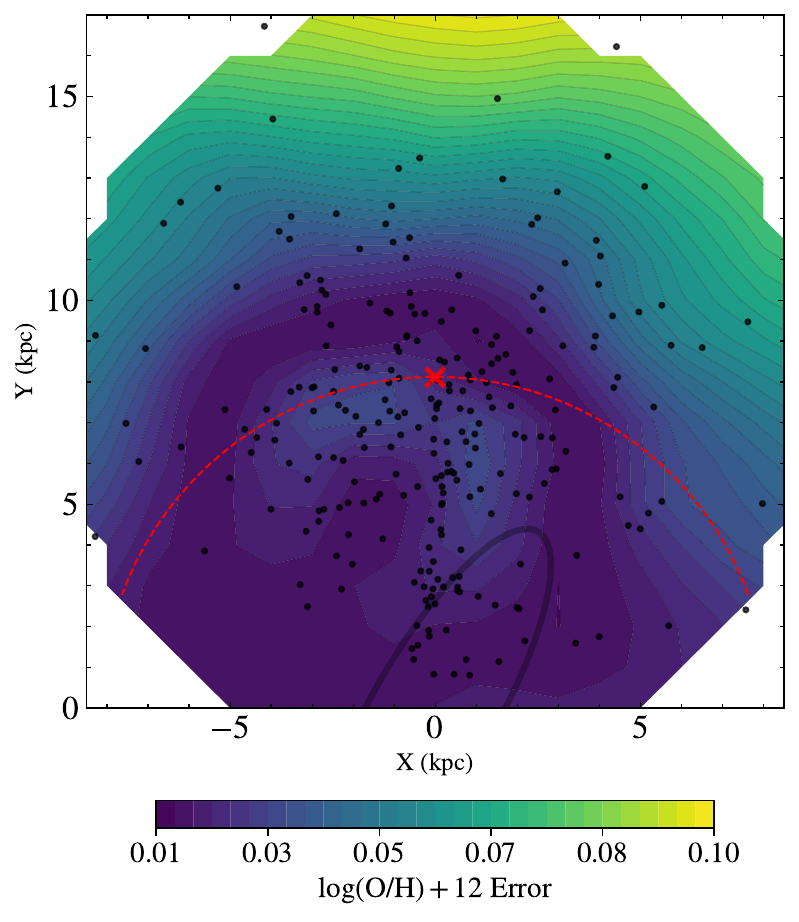}
\caption{Top: the $\log(\mbox{O}/\mbox{H}) + 12$ abundance map interpolated using the 2D universal kriging algorithm.
Bottom: the respective $\log(\mbox{O}/\mbox{H}) + 12$ error map. In both panels the red x marks the position of the Sun
and the red dashed semicircle the solar radius. The gray ellipse centered at the origin represents the position of the
Galactic bar after \citet{wegg15}. \label{fig:kriging_maps}} \end{figure}

The results are shown in Fig. \ref{fig:kriging_maps}, where in the top panel the O/H abundance map is presented in a
face-on view of the Galactic plane and in the bottom panel the respective error map in the O/H abundances. The
differences in the interpolated chemical abundances in the O/H map are not very large, but the map reveals an azimuthal
asymmetry in the distribution. The abundances are higher near the solar circle and near the Galactic Center at positive
longitudes. The higher abundances near $X = 0$ and $Y=5$~kpc seem to coincide with end of the expected position of the
Galactic bar. This asymmetry in the abundance distribution was also noted by \citet{recio-blanco23} using stellar
chemical abundances from \textit{Gaia} DR3 data and may be related to radial gas flows induced by the Galactic bar from
the outer Lindblad resonance (OLR) to the bar corotation radius, as predicted by chemical evolution models
\citep{cavichia14, kubryk15}. In the O/H map it is also possible to detect the radial O/H abundance gradient, as beyond
the solar circle the abundances drop continuously. The solar circle appears to be a transition region in the pattern of
chemical abundance distribution: inside the solar circle the abundances are higher than in the outside region, creating
a bimodal pattern of chemical abundance distribution. This characteristic is consistent with the fits obtained in the
one-break and two-break models presented in Fig. \ref{fig:grads_breaks}, where breaks in the radial distribution are
detected near the solar circle.

\subsection{Thin and thick disks O/H radial gradients and the origin of the metallicity breaks}

\begin{figure} \includegraphics[width = 1.0\columnwidth]{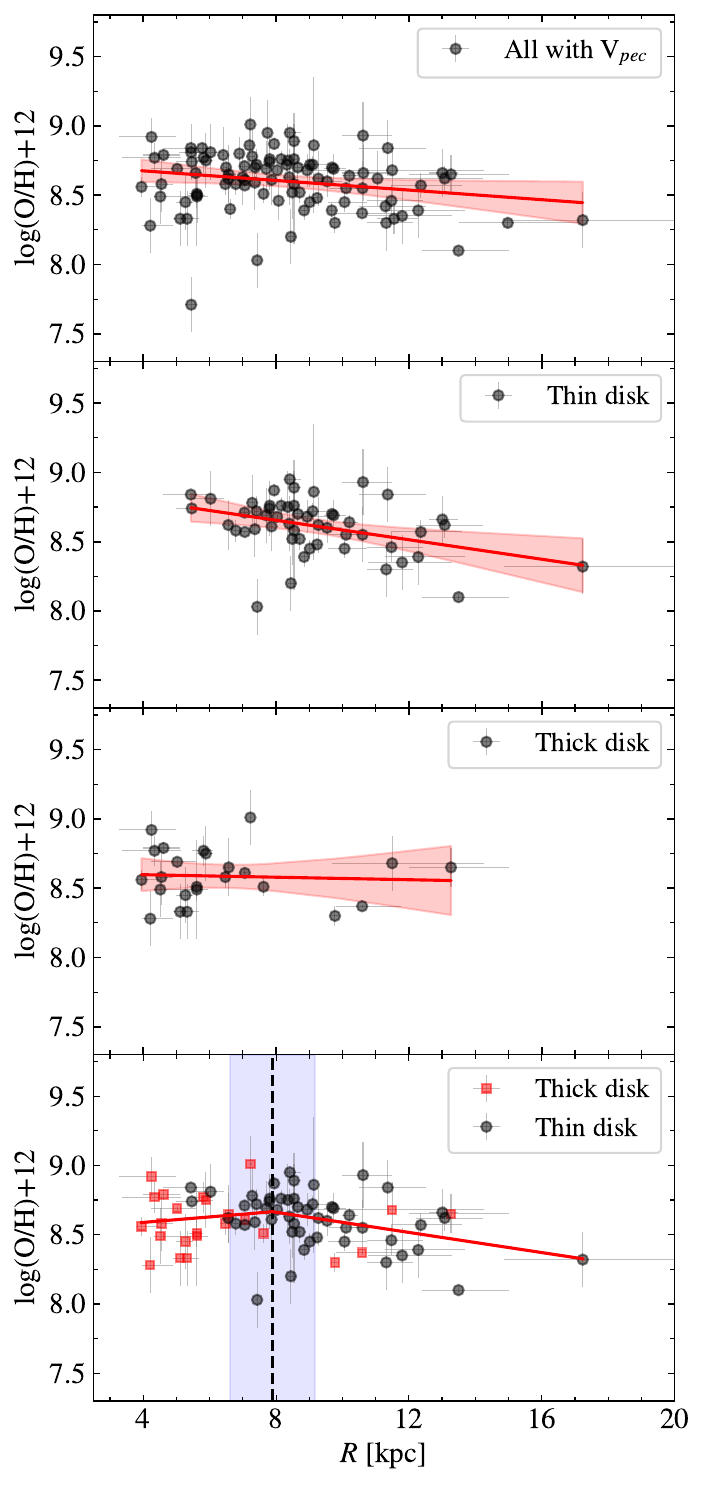} \caption{Radial abundance gradient for oxygen
including all disk PNe with calculated $V_\mathrm{pec}$, thin and thick disks PNe, as indicated. In each panel the red
line is the single linear fit and the shaded region the 95\% confidence interval. In the last panel a piecewise linear
regression with one break is fitted to the combined thin- and thick-disk data. \label{fig:disk_grads_thin_thick}}
\end{figure}

To further investigate O/H radial gradient in the Galaxy, we separate the disk PNe in the Toomre diagram (see Fig.
\ref{fig:toomre_diag}) into those pertaining to the thin disk ($V_{\mathrm{pec}} < 50$\,km/s) and to the thick disk 
($70 < V_{\mathrm{pec}} < 220 $\,km/s), as suggested by \citet{bensby14}. The top panel of Fig.
\ref{fig:disk_grads_thin_thick} shows the O/H radial gradient including all objects in our sample with measured
$V_{\mathrm{pec}}$. In the second panel, the O/H radial gradient for PNe is depicted from the thin disk, and in the
third panel the radial gradient for the thick disk PNe. Table \ref{tab:grads_thin_thick} lists the following results of
the performed linear fits: the sample name; the number of objects in each sample; slope and respective error; and
intercept and respective error. The thin disk O/H radial gradient obtained in this work is steeper than that of the
thick disk and more in agreement with that of the \Hii regions \citep[e.g.][]{esteban18,esteban17}. In particular,
\citet{arellano-cordova20} obtained the slope and intercept of $-0.037 \pm 0.009$~dex/kpc and $8.78 \pm 0.08$ dex,
respectively, for disk \Hii regions using distances based on \textit{Gaia} DR2 data, which are in excellent agreement
with the one obtained here for the thin disk. Additionally, the dispersion of the data around the linear relation for
thin-disk PNe is lower than considering all-disk PNe as in Fig. \ref{fig:grads_breaks}. However, the dispersion
observed for thin-disk PNe remains larger than the oxygen gradients recently derived for disk \Hii regions using large
aperture telescopes \citep[e.g.][]{arellano-cordova20,mendez-delgado22}. This discrepancy may be attributed to spectral
quality, the chemodynamical evolution of the Galaxy, the nucleosynthesis of oxygen within PN progenitor stars, or
heterogeneous chemical abundance samples. It is likely that a combination of these effects is occurring.

\begin{deluxetable}{lccc} \setlength{\tabcolsep}{10pt} \tabletypesize{\small} \tablewidth{\columnwidth}
\tablecaption{Radial O/H gradients for PNe with $V_{\mathrm{pec}}$. \label{tab:grads_thin_thick}} \tablehead{
\colhead{Sample} & \colhead{$N$} & \colhead{Slope} & \colhead{Intercept}\\ \colhead{} & \colhead{} & \colhead{(dex/kpc)}
& \colhead{(dex)}} \colnumbers \startdata All  & 99 & $-0.017 \pm 0.008$ & $8.74\pm0.07$ \\ Thin & 53 & $-0.035\pm0.012$
& $8.94\pm0.11$\\ Thick & 24 & $-0.005\pm0.017$ & $8.62\pm0.11$\\ \enddata \tablecomments{In this table the radial O/H
gradients are calculated using only disk PNe with measured peculiar velocities $V_{\mathrm{pec}}$.} \end{deluxetable}

The selection of PNe with lower values of $V_{\mathrm{pec}}$ implies an age criterion, since younger objects tend to
follow the Galactic rotation curve closer \citep{maciel11}. The dispersion of the data around the linear relation seen
in the top panel of Fig. \ref{fig:grads_breaks} may be an effect of the chemical evolution of the Galaxy, as probably
the range of ages of the progenitors of the PNe are much higher, increasing the dispersion in the abundances.
Additionally, the radial migration of the progenitor stars may also affect the dispersion, contributing to mixing PNe
from different radii in the Galaxy and, therefore, different chemical abundances. However, separating both effects is
out of the scope of this paper. The O/H radial gradient for thick disk PNe is shallower than that of thin disk, and
there is a large number of PNe within the solar radius ($R < 8$~kpc). Such a feature was also observed by
\citetalias{bucciarelli23}, however, they use a chemical abundance criterion to classify older and younger PNe, which
relies on models of stellar evolution and is subject to the related uncertainties. As can be seen in their figure 8, the
younger sample has a higher scatter of the data around the linear relation, and the gradient is shallower than the thin
disk PNe obtained in this work. In the fourth panel of Fig. \ref{fig:disk_grads_thin_thick} the combined thin and thick
disks data are plotted and visually it is possible to observe a flattening of the gradient inside the solar region. We
fitted piecewise linear functions to the O/H gradients of the thin disk, thick disk, and the combined thin and thick
disk populations, allowing for one or two breaks. However, the fit for a one-break model converged only for the combined
sample. The fit yields a break at $7.89\pm1.3$~kpc, as indicated by the vertical dashed line and the shaded region in
the figure. Consequently, the breaks in the metallicity radial gradients observed in this work may result from the
superposition of distinct stellar populations associated with the thin and thick disks.

\section{Discussion and conclusions \label{sec:conclusions}}

A long-standing discussion regarding the radial abundance gradient from PNe concerns the constancy of its slope across
the Galactic disk. Previous results suggest either a flattening of the oxygen gradient at large Galactocentric distances
\citep[e.g.,][]{maciel09,stanghellini18} or, conversely, a steepening \citep[e.g.,][]{henry10}. Part of this
disagreement arises from the different methods used to estimate the PN distances. To address this issue, in this work,
we provide a recalibration of the H$\alpha$ surface brightness--radius relation from \citetalias{frew16}, based on
\textit{Gaia} DR3 data, to derive distances for 1,130 Galactic PNe. In addition, Bayesian distances are provided for a
subsample of 415 PNe with available \textit{Gaia} parallaxes. Using these distances we derived the O/H radial gradient
for a sample of 231 disk PNe.

We present the two-dimensional distribution of PNe in the Galactic plane and compare it with the expected locations of
the spiral arms and the Galactic bar. The distribution shown in Fig.~\ref{fig:mw_spiral_arms} and the discussion
presented in Section \ref{appendix_b} in the Appendix support the reliability of the Bayesian distances derived in this
work, indicating that they are among the most robust PNe distances currently available to investigate the radial
chemical abundance gradients in the Milky Way.

To fit the radial metallicity gradient for disk PNe we considered three possibilities: a single linear fit, a segmented
fit with one break, and a segmented fit with two breaks. The single linear fit resulted in a slope of
$-0.020\pm0.006$~dex/kpc, in agreement with the most recent O/H radial gradient from PNe
\citep{bucciarelli23,stanghellini18}. The O/H radial gradient derived for the disk PNe sample using segmented fit
exhibits a significant break near the solar region, as revealed by segmented linear models. These fits, incorporating
one or two radial breaks, indicate a flatter, or even slightly positive, gradient in the inner disk for $R \lessapprox
8$\,kpc. Beyond this radius, the gradient transitions to a steeper and more negative slope compared to a single linear
fit. This potential change of slope in the radial gradient at $R \approx 8$\,kpc deserves further investigation.
According to the AIC, the selection of the best model  between a segmented fit with one break and a segmented
fit with two breaks, is inconclusive, implying that the two are plausible given the current data. However, given the AIC
difference, the linear model fit is notably weaker than the one-break model.

The results found in this work are generally consistent with previous findings in the literature. The flattening
observed in Cepheids \citep{martin15,andrievsky16} and in red giants \citep{hayden14} appears to occur at smaller
Galactocentric radii ($R \sim 5 \mbox{--} 6$\,kpc). More recently, \citet{andrievsky26} derived iron, sulfur and oxygen
abundance gradients for Cepheid stars using distances from \textit{Gaia} DR3. They find evidence for a change in the
slope of the radial abundance distribution starting at 10 kpc and a flattening for Galactocentric distances larger than
14 kpc. \citet{malhan25} analyzing  metal-poor stars data from APOGEE DR17 and \textit{Gaia} DR3 find that for 9
different element abundances exhibit a drastic transition in their distribution near the Solar radius.
\citet{stanghellini18} proposed that the O/H radial distribution for disk PNe may be best described by a step function
near $R\sim10$~kpc rather than a continuous gradient. Their results are similar to those presented here, but with
important differences. We also find that the O/H gradient is flatter, or even positive, in the inner Galaxy; however,
the transition region is closer to the Sun. Another significant difference concerns the gradient in the outer Galaxy,
which remains negative and is steeper in both the one-break and two-break models than their suggested gradient. It is
also worth noting that \citet{stanghellini18} derived their results by fitting the radial O/H distribution by imposing
the step function manually, whereas our analysis is based on well-defined statistical criteria and a robust and
automatic fitting procedure. Nonetheless, other studies--primarily focusing on \Hii regions -- have found no
evidence of metallicity breaks \citep[e.g.][]{esteban17, arellano-cordova20, mendez-delgado22, martinez-hernandez26}.
The O/H gradient flattening for $R < 8$\,kpc reported by \citet{esteban18} was not supported by later works from the
same group. \citet{mendez-delgado22} demonstrate that this flattening is likely an artifact of underestimated
Galactocentric distances for certain objects.

Although statistically we cannot favor a single model, our results are qualitatively consistent with a change of slope
in the O/H radial gradient. The existence of two distinct gradient slopes is often attributed to variations in gas
infall and star formation rates between the inner and outer disk. One possible explanation for the change in slope is
the quenching of star formation efficiency caused by the dynamical action of the Galactic bar \citep[see][for a
discussion]{esteban18}. Simulations show that the non-axisymmetric gravitational potential of the bar drives gas inflows
toward the central regions of galaxies \citep[e.g.][ and references therein]{fragkoudi16}, and observations confirm
bar-driven gas transport \citep[e.g.][and references therein]{lopez-coba22}. According to the chemical evolution model
(CEM) of the Milky Way by \citet{cavichia14}, the bar induces radial gas flows within the corotation radius, increasing
gas density while diluting abundances and simultaneously enhancing the star formation rate. \citet{carigi19} proposed a
CEM in which the flattening of the oxygen gradient in the inner Galaxy arises from inside-out quenching of the star
formation history. The origin of this quenching may be gas flows toward the Galactic Center induced by the bar.

\citet{reddy16} argued that the change in slope may instead reflect a mixture of stellar populations at different
vertical distances from the Galactic plane: clusters close to the mid-plane trace a steeper radial metallicity gradient,
whereas those located farther away trace a shallower one. Similar results are obtained for stars located in the
thin and thick disks \citep{vickers21}. These results are consistent with the ones obtained in this work for PNe when
comparing the O/H radial gradients of the thin and thick disks, with the thin-disk PNe showing a steeper gradient. The
piecewise regression analysis presented in Fig. \ref{fig:disk_grads_thin_thick} further supports this interpretation.
Upon analyzing the kinematically separated thin and thick disk PNe, the fitted model indicates a break in the
metallicity gradient at $7.89\pm1.3$~kpc for the combined thin and thick disk populations. Since the thick disk exhibits
a flatter metallicity gradient compared to the thin disk, the superposition of both populations generates an apparent
break in the metallicity gradient near the solar neighborhood. This may also explain why recent studies focused
on \Hii regions do not find evidence of metallicity breaks in the Milky Way disk, as these objects trace the thin disk
metallicity gradient.

The formation of the thick disk has been linked to radial migration of stars as a consequence of the spiral structure
\citep{Schonrich09,schonrich09b}, producing flat or positive thick-disk radial metallicity gradients. The results
obtained in this work for thin and thick disks are consistent with the chemodynamical models of
\citet{minchev13,minchev14}, where stars younger than 2 Gyr exhibit a steep gradient and a flat trend for stars older
than 2 Gyr throughout the radial range 5--16 kpc of the disk. The flattening of the radial gradient in their simulations
is a result of the radial mixing of stars and is an interpretation for the thick disk formation, as it emerges naturally
from stars migrating from the inner disk very early on due to strong merger activity, followed by additional radial
migration driven by the bar and spirals at later times. \citet{kubryk15} using a CEM with radial motions of both gas and
stars obtained similar conclusions, as the metallicity profiles of the thick disk -- assumed as the oldest stars -- are
flatter, especially at the inner disk, due to the radial motions induced by the Galactic bar.

Another possibility is that the break in the O/H radial gradient is related to the corotation resonance of the spiral
pattern. It has been suggested that the Galaxy undergoes a bimodal chemical evolution, since gas kinematics has opposite
directions at the corotation radius. Using a homogeneous open cluster sample with \textit{Gaia} DR2 data, \citet{dias19}
determined the corotation radius to be located at $8.51 \pm 0.64$\,kpc, close to the solar orbit. This value agrees,
within the uncertainties, with the break in the O/H gradient derived from PNe in this work. Breaks in the
abundance gradient near the corotation radius have been reported by many authors using different tracers
\citep[e.g.,][]{carney05,lepine11,genovali14,reddy16,stanghellini18,monteiro21,magrini23,dasilva23,yang25,andrievsky26}.
Breaks in the O/H radial gradient have also been reported in other spiral galaxies
\citep{sanchez14,sanchez-menguiano16,sanchez-menguiano18,easeman22,pilyugin24,cardoso25,perez-diaz25,pilyugin26}. In
particular, \citet{sanchez-menguiano18} analyzed the O/H radial distribution of \Hii regions in a sample of 102 spiral
galaxies observed with VLT/MUSE and detected a characteristic break at $\sim 1.5$\,\textit{R}\textsubscript{e}, although
with significant variation among individual galaxies. Adopting an effective radius of $R_{\rm e} = 5.85$\,kpc for the
Milky Way \citep{molla19a}, 1.5\,\textit{R}\textsubscript{e} corresponds to $\sim 8.8$\,kpc, which is consistent,
within the uncertainties, with the position of the breaks suggested by both the one-break and two-break models
derived in this work.

To investigate the O/H azimuthal distribution in the Galactic plane, we applied a kriging interpolation algorithm. The
results reveal an azimuthal asymmetry, with higher abundances near the bar position at positive longitudes. However, the
abundance variations are modest and remain within the measurement uncertainties. In the O/H map it is also possible to
note the bimodal abundance pattern for the inner and outer solar regions.  The azimuthal metallicity structure of the
Milky Way has also been studied using \ion{H}{0ii} regions
\citep[e.g.,][]{balser11,balser15,wenger19,martinez-hernandez26}. Some of these works report differences of up
to a factor of two in the O/H radial gradient depending on the azimuthal angle considered. Similarly to our results, the
O/H maps from \citet{balser15,wenger19} show enhanced abundances near the Galactic bar at positive longitudes. However,
their analysis lacks \Hii regions in the 3rd and 4th Galactic quadrants, preventing a direct comparison with the results
obtained here. On the contrary, \citet{martinez-hernandez26} did not find clear evidence of such enhanced
abundances. However, similarly to the other studies, their analysis also lacks observations in 3rd and 4th quadrants.

The findings obtained in this work emphasize the complex interplay between Galactic dynamics, star formation, and
chemical evolution, highlighting the importance of considering both radial and azimuthal variations in abundance
studies. Future work combining larger, homogeneous samples with improved distance and abundance determinations will be
essential to confirm the presence and origin of the observed gradient break and to refine models of Galactic chemical
evolution.

\begin{acknowledgments} We thank the anonymous referee for a careful and constructive report that significantly
improved the quality and clarity of this paper. This work has made use of the computing facilities available at the
Laboratory of Computational Astrophysics of the Universidade Federal de Itajub\'a (LAC-UNIFEI). The LAC-UNIFEI is
maintained with grants from CAPES, CNPq and FAPEMIG. OC and MC have been funded through grant PID2022-136598NB-C33 by
MCIN/AEI/10.13039/501100011033 and by ``ERDF A way of making Europe''. This research has made use of the HASH PN
database at hashpn.space \citep{parker06}. This work has made use of data from the European Space Agency (ESA) mission
\textit{Gaia} (\url{https://www.cosmos.esa.int/gaia}), processed by the \textit{Gaia} Data Processing and Analysis
Consortium (DPAC, \url{https://www.cosmos.esa.int/web/gaia/dpac/consortium}). Funding for the DPAC has been provided by
national institutions, in particular the institutions participating in the \textit{Gaia} Multilateral Agreement.

Software: Astropy \citep{astropy13,astropy18,astropy22}, Matplotlib (Hunter 2007), NumPy \& SciPy \citep{scipy20},
PyKrige \citep{murphy14,benjamin_murphy25}, Python (\url{https:// www.python.org/}). \end{acknowledgments}

\appendix

\section{Monte Carlo linear fit procedure \label{appendix_a}}

\begin{figure} \plotone{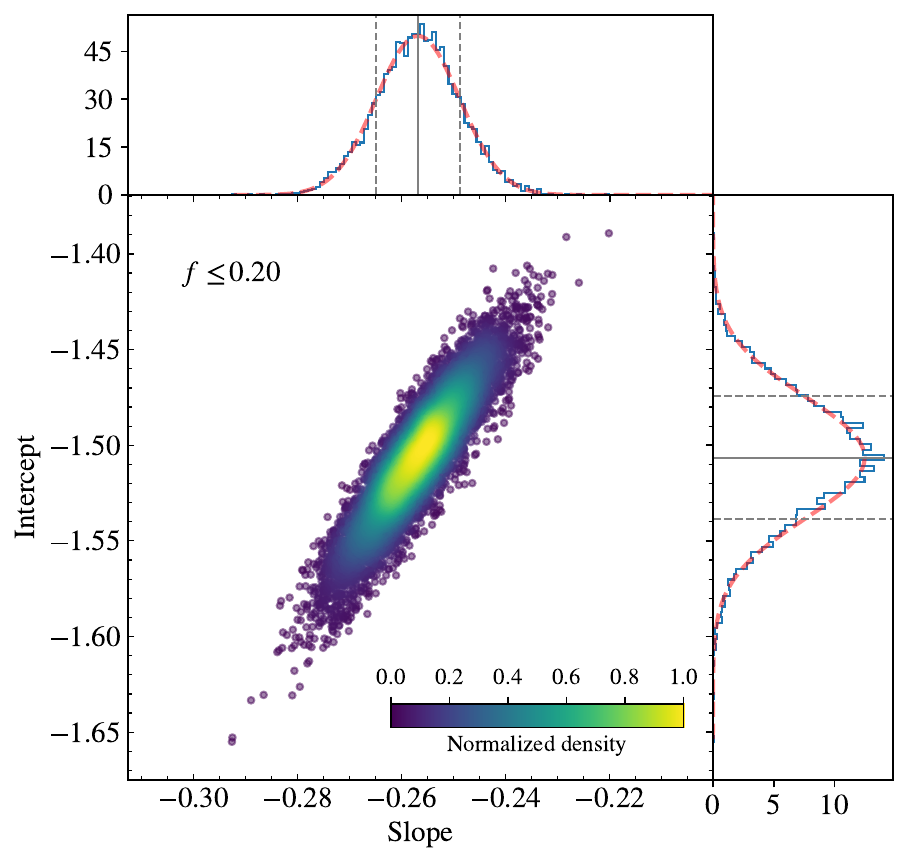} \caption{Distributions of slope and intercept of the linear fits using an MC
simulation for the $f$ cases analyzed in Fig. \ref{fig:r_sha_frac_erro}. In each axis the histograms of the
distributions are fitted with a Gaussian, showed as the red dashed lines. The mean and the 68\% confidence interval of
the Gaussian are shown as the vertical continuous and dashed lines, respectively. The complete figure set (4
images) is available in the online journal.} \label{fig:mc_linear_fit} \end{figure}

In this paper we present the recalibration of the H$\alpha$ surface brightness--radius relation with a large number of
PNe with accurate astrometric parallaxes are available from \textit{Gaia} DR3. The linear fits presented in Figs.
\ref{fig:r_sha_frac_erro} and  \ref{fig:r_sha_optically} are obtained using 10,000 MC realizations by varying the
angular diameters and the H$\alpha$ surface brightness within their uncertainties. For each MC realization, an
OLS regression is performed and, this way, generating distributions of the slopes and intercepts. Each
distribution is fitted with a Gaussian and the values are adopted as the mean and standard deviation. Fig.
\ref{fig:mc_linear_fit} shows the results for the $f \le 0.20$ case as an example and this procedure was performed for
all the $f$ cases analyzed in this work and the figures associated to the other cases are available in the
online journal. The resulting mean and the 68\% confidence interval of the slope and intercept obtained from this
procedure are depicted in the histograms and correspond to $-0.257 \pm 0.008$ and $-1.50\pm0.03$, respectively.

\section{Bayesian and statistical distance methods comparison \label{appendix_b}}

\begin{figure*}[!ht] \centering \includegraphics[width=0.9\linewidth]{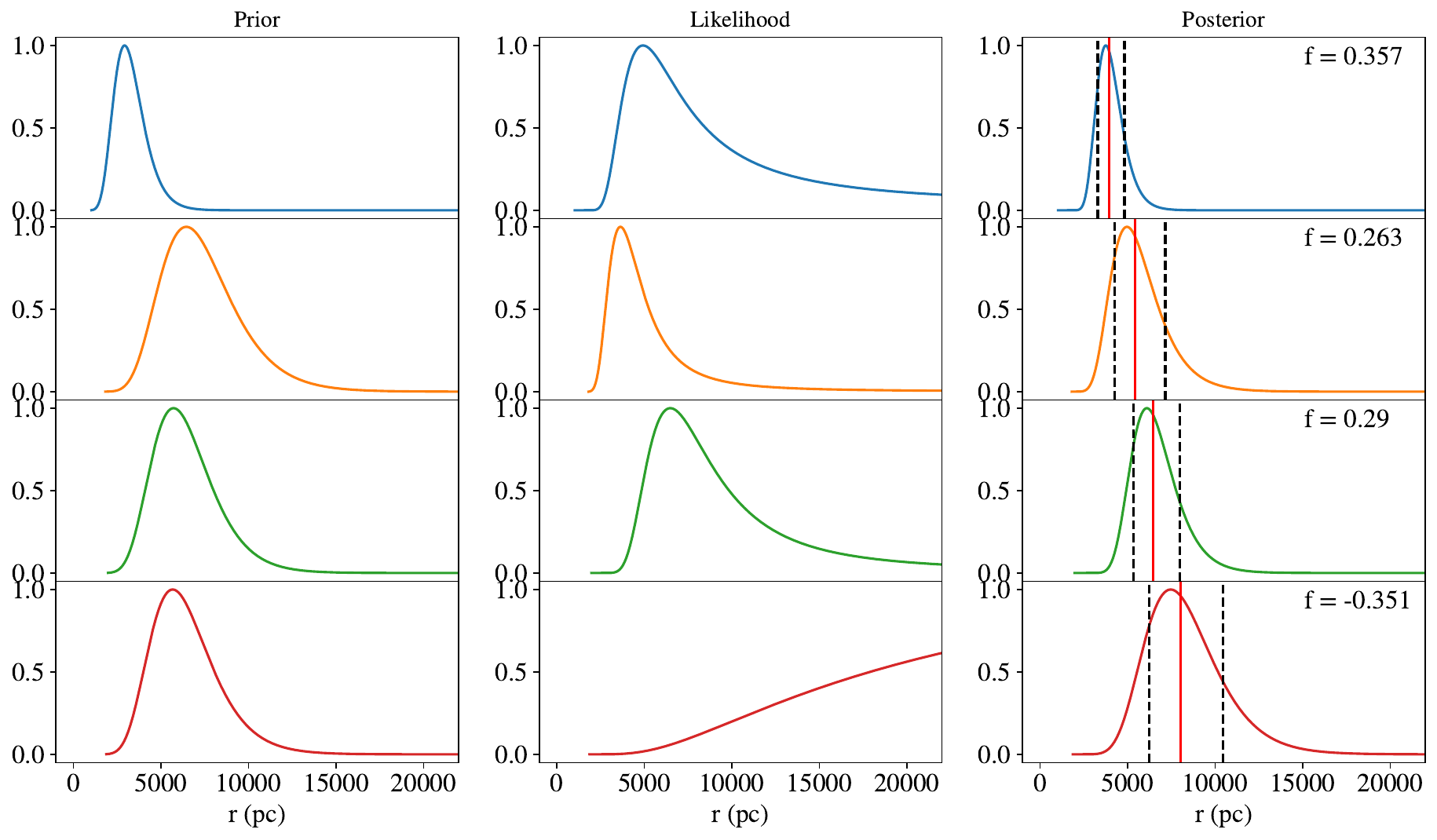} \caption{Left panels: normalized priors
distributions using statistical distances from the H$\alpha$ surface brightness-- radius relation as recalibrated in
this work. Middle panels: the normalized likelihoods computed from the parallaxes. Right panels: the respective
normalized posterior distributions. The vertical continuous line represent the median value of the posteriors and the
vertical dashed lines the 16th and 84th percentiles. The sources have fractional parallax uncertainties $f$ as labeled
in the right panels. \label{fig:priors_posteriors}} \end{figure*}

The recalibration of the H$\alpha$ surface brightness--radius relation is used in Section \ref{sec:bayes} to obtain more
reliable distances for a smaller sample of 415 PNe using Bayesian statistics, adopting the statistical distances and
their uncertainties as a prior and the \textit{Gaia} parallax \parallax\ and its uncertainty \parallaxsd\ as the
likelihood \citep{bailer-jones21, chornay21}. Because many of our posteriors are not well approximated by Gaussian
distributions, we compute the posterior distance distribution on a fixed grid with 0.01 pc steps between 0 and 50 kpc,
and the distributions are normalized by the sum of values over the domain of the prior. To estimate the distances and
the uncertainties, we report the median and the 16th and 84th percentiles of the posterior distribution. The results for
four PNe with different $f$ values are displayed in Fig. \ref{fig:priors_posteriors} as examples of the resulting
distributions. In the first, second, and third rows, where $f$ is positive, the posterior distributions are narrower
than the prior distributions. In the fourth row, where $f$ is negative, the posterior distribution tends toward the
prior distribution.

In order to assess the improvement in the calculation of Bayesian distances over the statistical distances alone, we
uniformly generate random data for 5000 PNe with true physical radii $R_{\mathrm{true}}$ between 0.03 and 1 pc and true
angular radii $\theta_{\mathrm{true}}$ between 5 and 100 arcsec. We then calculate the true values
$\log(S_{\textrm{H}\alpha})_{\mathrm{true}}$ by assuming the calibration of the ${\log(S_{{\rm H}\alpha}})$ --
$\log(R_{\rm pc})$ relation given by equation \ref{eq:halpha_radius} and an intrinsic dispersion of 27\% over the
relation calculated using a maximum likelihood estimation method. The observational values
$\log(S_{H\alpha})_{\mathrm{obs}}$ are calculated by adding Gaussian-distributed errors, assuming an average uncertainty
of 0.17 dex derived from the uncertainties provided by \citetalias{frew16}. The same procedure is applied to the angular
radii, but in this case assuming an average relative error of 25\%.

Using the observed values $\log(S_{\textrm{H}\alpha})_{\mathrm{obs}}$ and $\theta_{\mathrm{obs}}$ and applying equation
\ref{eq:halpha_radius}, we obtain the statistical distances based on $S_{\textrm{H}\alpha}$. The true distances
$d_{\mathrm{true}}$ are calculated from $R_{\mathrm{true}}$ and $\theta_{\mathrm{true}}$, and from $d_{\mathrm{true}}$
we obtain the true parallaxes. The Bayesian distances are then calculated by adding Gaussian-distributed errors to the
parallaxes and applying Bayes’ theorem, using the statistical distances from the H$\alpha$ surface brightness--radius
relation as priors. In the analysis we consider parallax fractional errors $f$ between 0.1 and 0.9 distributed in steps
of 0.1.

\begin{figure*}[!ht] \centering \includegraphics[width=1.0\columnwidth]{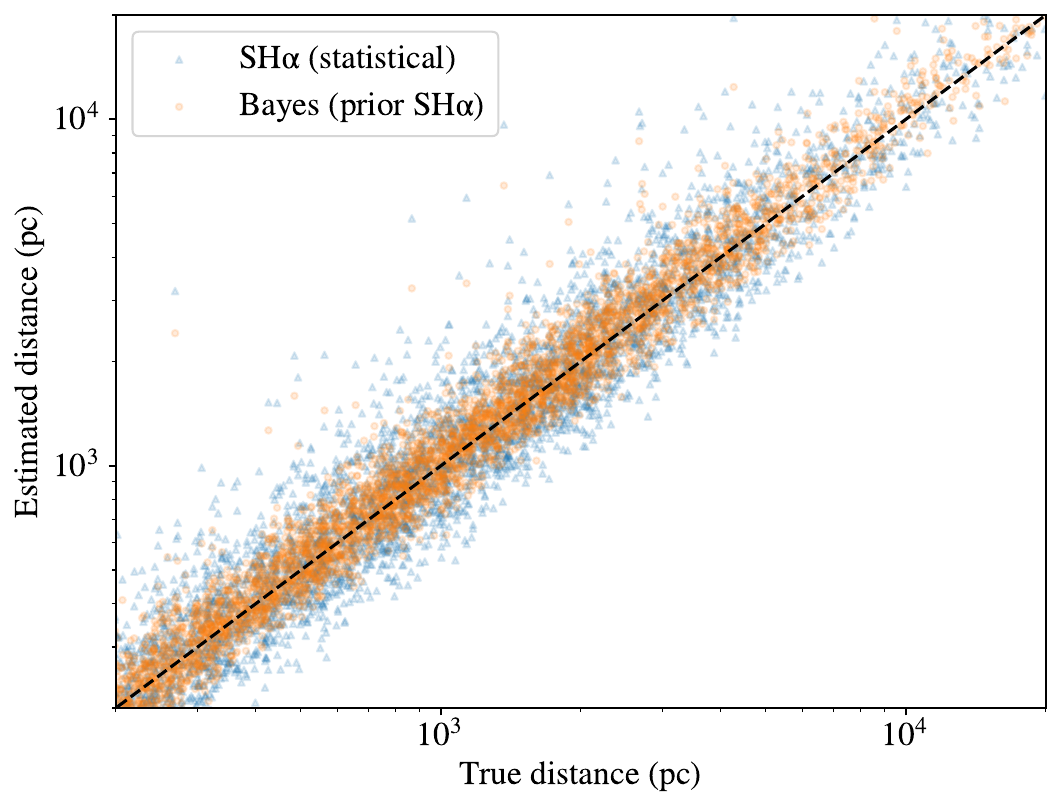}
\includegraphics[width=1.0\columnwidth]{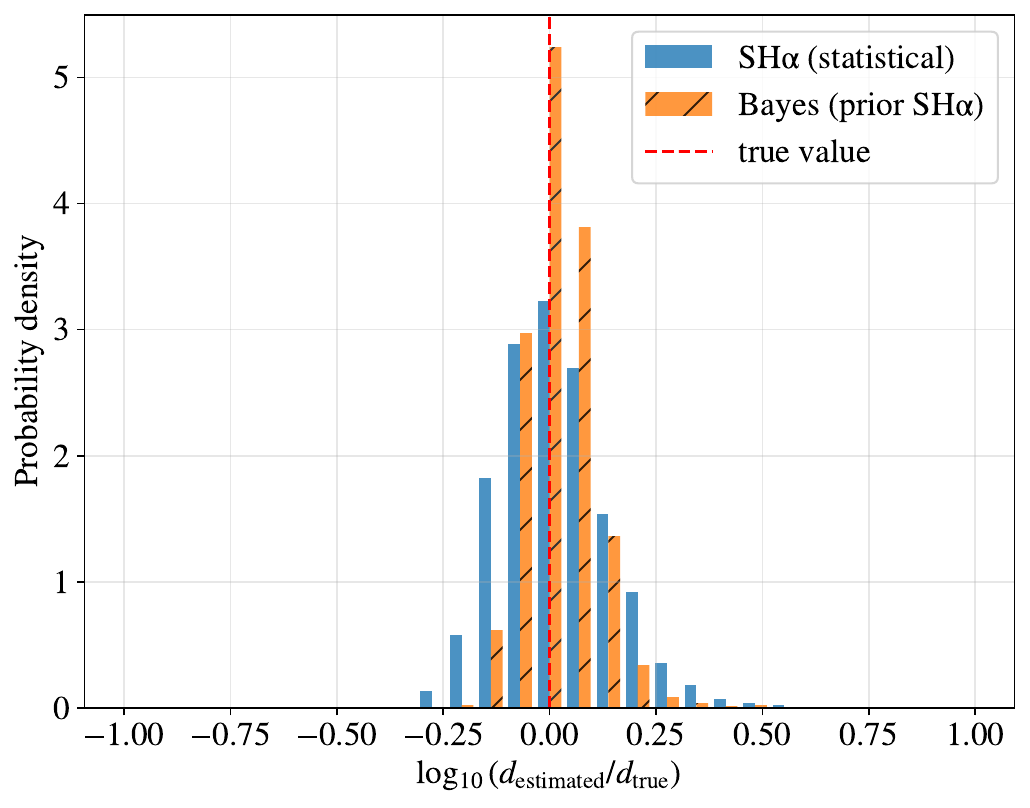} \includegraphics[width=1.0\columnwidth]{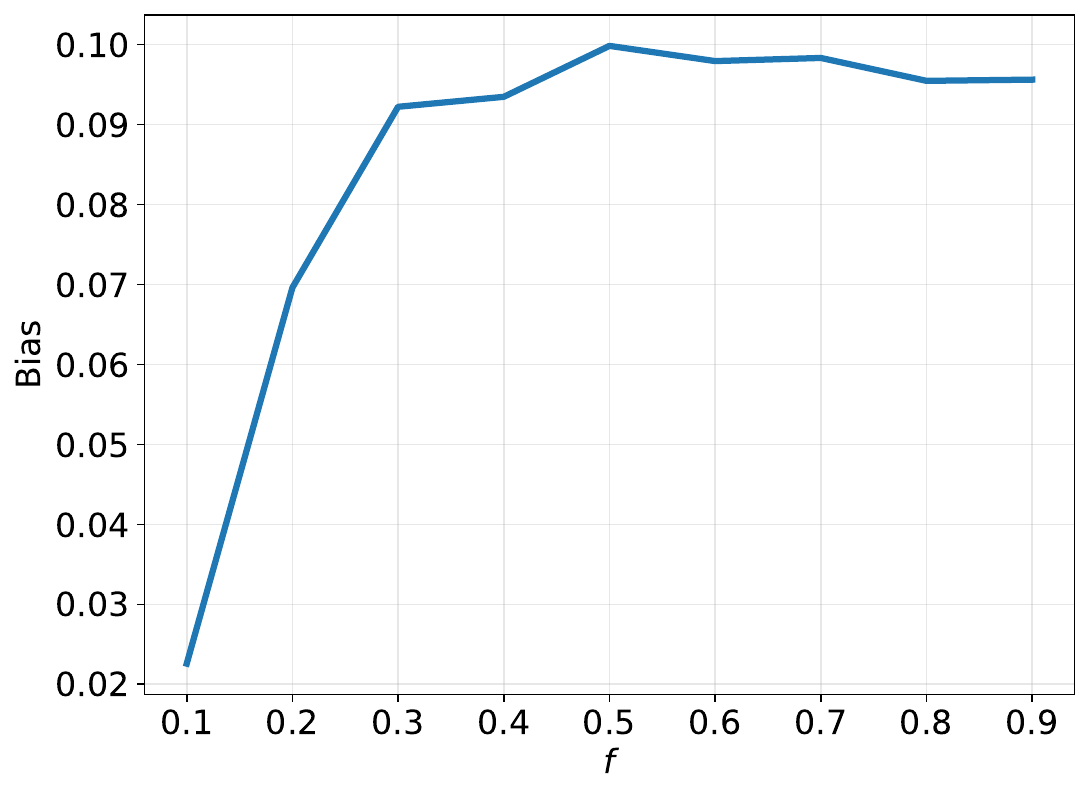}
\includegraphics[width=1.0\columnwidth]{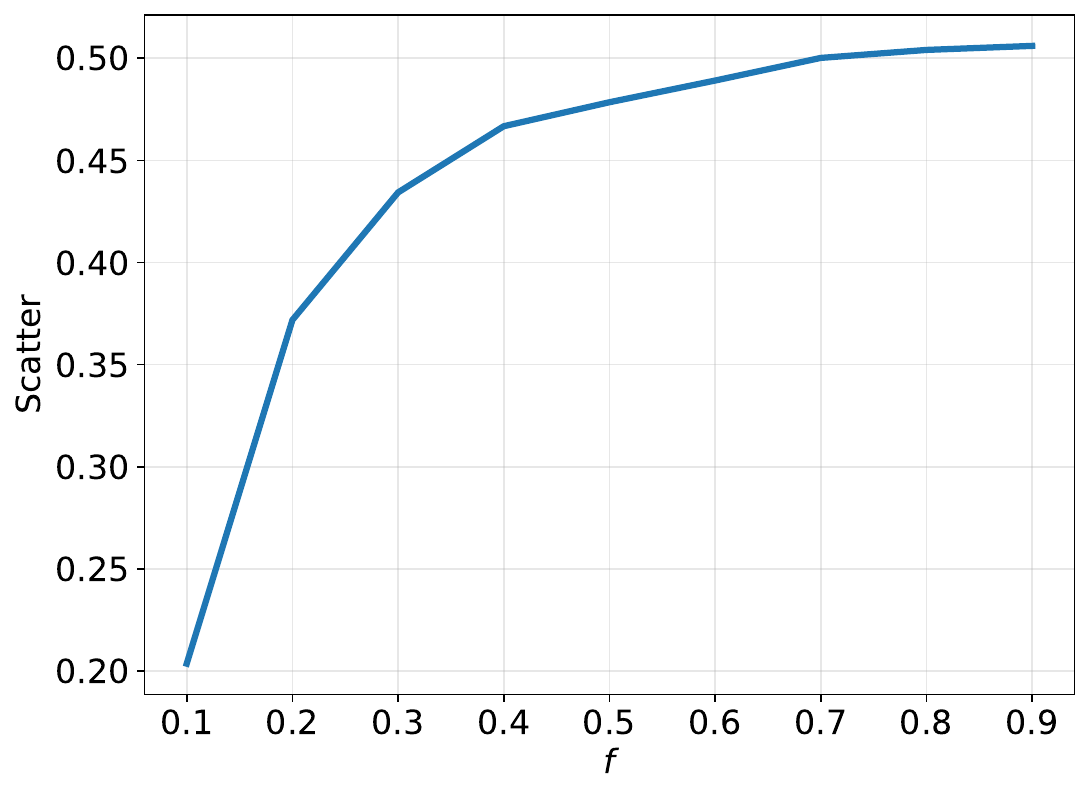} \caption{Top left: comparison between the simulated distances derived
from the $S_{\textrm{H}\alpha}$ statistical distances and the Bayesian distances for the case $f = 0.2$. The dashed line
corresponds to the equality. Top right: histograms of the simulated distances comparing the estimated and the true
distances for the case $f = 0.2$. The vertical dashed line is the true value. Bottom left and right: the bias and the
scatter of the distributions as a function of the fractional parallax uncertainty $f$, respectively.
\label{fig:estimated_true_dist}} \end{figure*}

The top panels of Fig. \ref{fig:estimated_true_dist} show the results for the case $f = 0.2$, comparing the estimated
distances with the true distances derived from the $S_{\textrm{H}\alpha}$ statistical distances and the Bayesian
distances. Defining the relative error $(d_{\mathrm{estimated}} - d_{\mathrm{true}}) / d_{\mathrm{true}}$, the bottom
panels show the bias and the scatter of the distributions, calculated for each value of $f$, as the mean and the
standard deviation of the relative error, respectively. For $f < 0.5$, both the scatter and the bias of the Bayesian
distances are lower than those of the $S_{\textrm{H}\alpha}$ statistical distances. For $f > 0.5$, the Bayesian values
tend towards the $S_{\textrm{H}\alpha}$ statistical bias and scatter values of 0.094 and 0.552, respectively, and the
two methods become equivalent. In the PNe sample with measured \textit{Gaia} parallaxes, there are 234 PNe
(56\%) with fractional parallax uncertainties in the range $0 < f < 0.5$; therefore, for these PNe, the Bayesian
distances are expected to be more accurate than the $S_{\textrm{H}\alpha}$ statistical distances.

\section{Comparison with other distance scales \label{appendix_c}}

\begin{figure*} \centering \includegraphics[width=1.0\columnwidth]{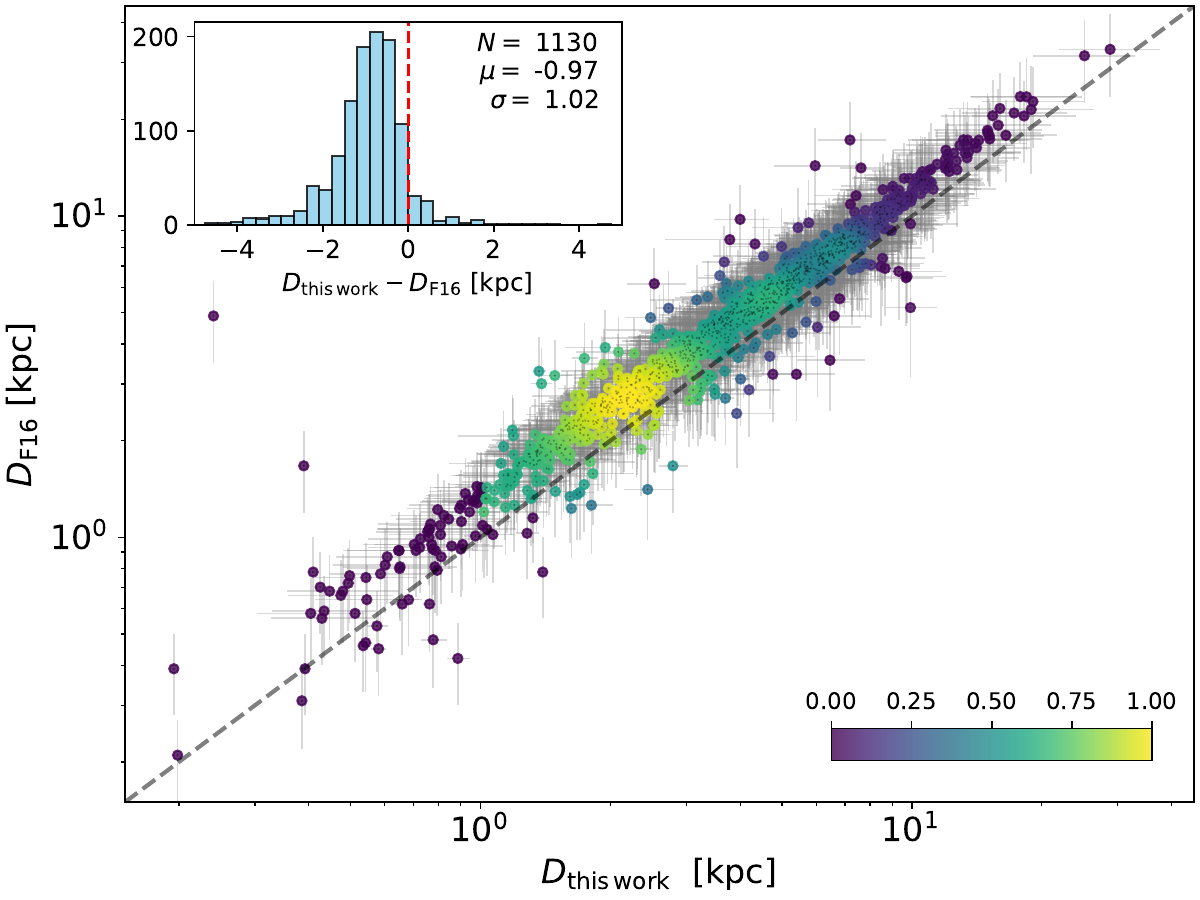}
\includegraphics[width=1.0\columnwidth]{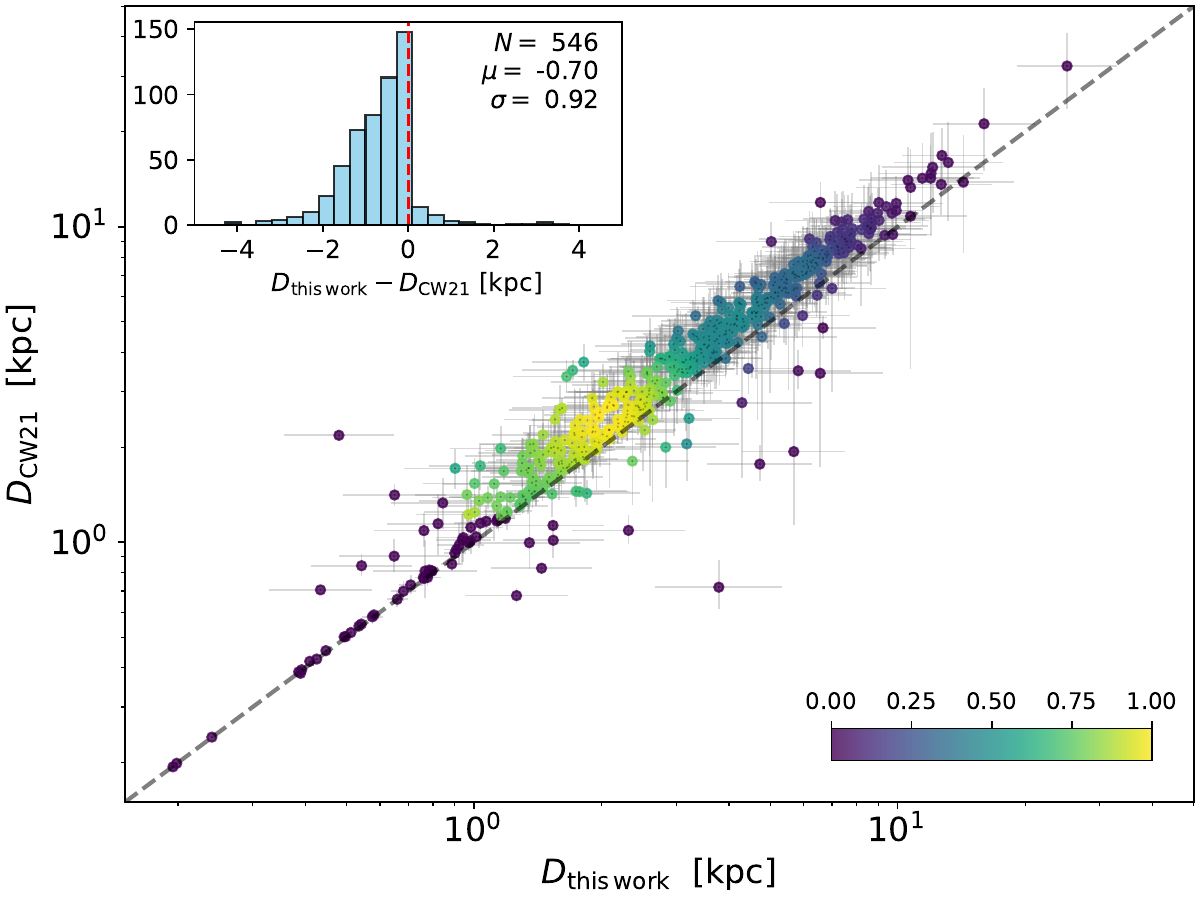} \includegraphics[width=1.0\columnwidth]{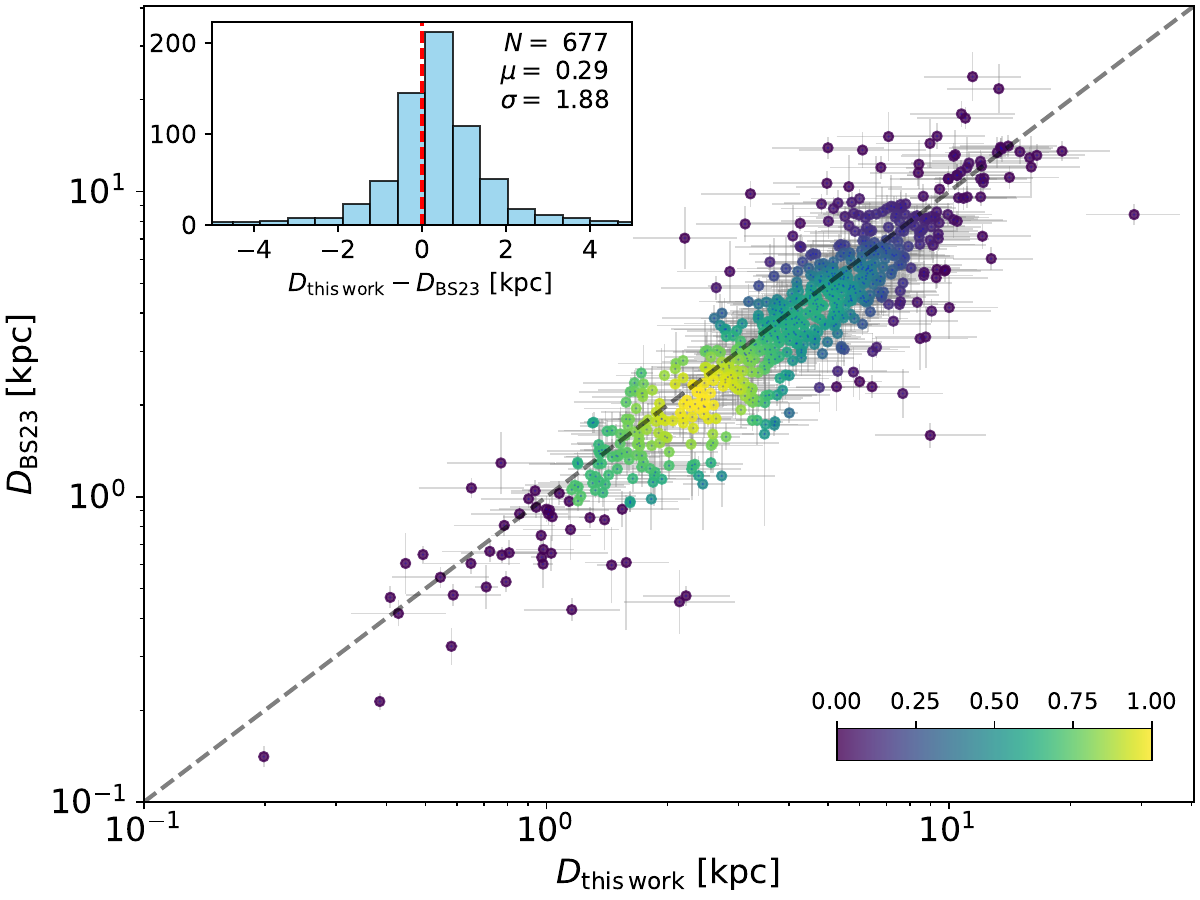}
\includegraphics[width=1.0\columnwidth]{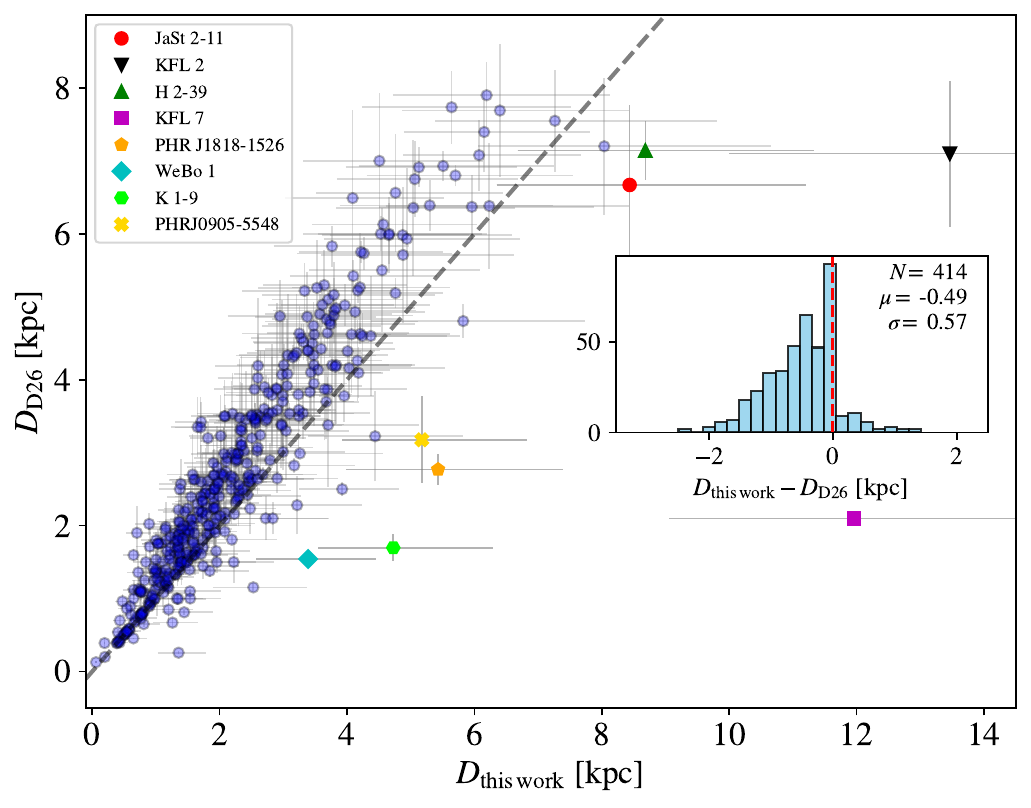}

\caption{Comparison of the distances obtained in this work ($x$-axis) and distances from \citetalias{frew16}
(top left), \citetalias{chornay21} (top right), \citetalias{bucciarelli23} (lower left) and  \citetalias{deng26} (lower
right). The dashed lines indicate equality. Panels comparing our distances with \citetalias{frew16},
\citetalias{chornay21} and \citetalias{bucciarelli23} are in a logarithmic scale for a better visualization. The insets
show the histograms of the residual distributions with the mean ($\mu$) and scatter ($\sigma$) and the vertical red
dashed lines mark the equality. The number of data $N$ in common between the samples is indicated in each inset. Some of
the largest differences with \citetalias{deng26} are highlighted with different symbols and colors as labeled with the
PNe names. \label{fig:dists_comparision}} \end{figure*}

The final distances adopted in this paper are Bayesian distances based on the $S_{\textrm{H}\alpha}$ statistical
distances and the \textit{Gaia} DR3 parallaxes. For objects without reliable parallaxes, the distance distribution was
obtained directly from the log-normal prior derived from the $\log R_{\rm pc}$\,--\,$\log S_{\rm H\alpha}$ relation. In
Fig. \ref{fig:dists_comparision} we provide a comparison between our final distances and the most recent and
comprehensive statistical distances based on $S_{\textrm{H}\alpha}$ and $S_{\textrm{H}\beta}$ calibrations and
also based on extinction measurements. The comparison with the statistical distances from \citetalias{frew16} and with
1,130 PNe in common shows that, on the average,  \citetalias{frew16} distances are 26\% larger than the distances from
this work. A high density of data can be seen around the average trend. The PNe from \citetalias{chornay21} are filtered
using their parameters RUWE $< 1.4$ and reliability scores $> 0.8$. This yields 546 PNe in common with our final sample.
\citetalias{chornay21} distances are on average 19\% higher than ours. Particularly, the difference is higher for larger
distances, where their use the $S_{\textrm{H}\alpha}$ statistical distances from \citetalias{frew16} as priors. The
scatter of points below the identity line may reflects misidentification of \textit{Gaia} DR3 sources by
\citetalias{chornay21}. In the bottom left panel the comparison is made with \citetalias{bucciarelli23} distances and
with 677 PNe in common. On the average, \citetalias{bucciarelli23} distances are 6\% lower than our distances, however
with a larger dispersion around this value. The dispersion can be attributed to different sources of the data used to
derive the distances. We note that the dispersion increases for larger distances, indicating a possible effect of the
uncertainties in the angular radii and line fluxes. Most of the angular radii from \citetalias{bucciarelli23} are from
\citet{acker92}, \citet{cahn92}, \citet{tylenda03} and \citet{stanghellini08}, while H$\beta$ fluxes are from
\citet{acker92} and \citet{cahn92}. On the other hand, the angular radii used in this work are preferentially adopted
from \citetalias{frew16}, supplemented by data from \citet{tylenda03}, \citet{stanghellini08}, and \citet{acker92}.
\citetalias{frew16} compiled angular dimensions for the brighter PNe from \citet{tylenda03} and \citet{ruffle04}.
Angular dimensions for most of the largest PNe have been determined by \citetalias{frew16}, based on available digital
broad-band red or H$\alpha +$[\ion{N}{0ii}] images primarily taken from the SuperCOSMOS H-alpha Survey
\citep[SHS,][]{parker05,frew14} and INT Photometric H-Alpha Survey \citep[IPHAS,][]{drew05}.  For compact PNe the images
are obtained from the \textit{Hubble Space Telescope} (\textit{HST}). The H$\alpha$ fluxes used in this work are
obtained from \citetalias{frew16}, which are a compilation of literature data from \citet{kohoutek91, dopita97,
wright05, frew13} for the brighter objects, or from \citet{frew08, frew14} for a few of the largest and most evolved
PNe.

Recently, \citet[][hereafter D26]{deng26} determined distances for a sample of 1066 PNe by refining
extinction--distance measurements using \(\textit{Gaia}\) DR3 data. In the bottom-right panel of Fig.
\ref{fig:dists_comparision}, we compare these values with the Bayesian distances derived in this work. For most of the
422 PNe common to both samples the agreement is good; however, \citetalias{deng26} distances are, on average, 22\%
larger than ours and significant discrepancies exist for some PNe, which we highlight in the figure using distinct
symbols and colors. The inset displays the residuals after excluding these outliers.

The central stars (CS) of PNe JaSt 2-11, KFL 2, KFL 7, PHR J1818-1526, WeBo 1, and K 1-9 were not detected in
our study and were likely misidentified by \citetalias{deng26}. For KFL 7, although the source \textit{Gaia} DR3
4063052668692883072 is located near the geometric center of the PN and matches the distance estimated by
\citetalias{deng26}, \textit{Gaia} DR3 reports an effective temperature of $T_{\rm eff} = 4800$\,K, which is too low for
a CSPN. Similarly, for PHR J1818-1526, the \citetalias{deng26} distance coincides with the \citet{bailer-jones21} value
for \textit{Gaia} DR3 4098119397132285568; however, this source has $T_{\rm eff} = 11741$\,K from \textit{Gaia} DR3 and
a very low reliability score (0.25) according to \citetalias{chornay21}.

PN WeBo 1 is large and asymmetric; its CS was identified by \citetalias{chornay21} as \textit{Gaia} DR3
465640807845756160. This source has \({\rm RUWE} > 1.4\), indicating a poor astrometric fit often associated with
orbital motion, photocenter motion, or multiplicity. Consequently, our distance for this PN is derived solely from the
$\log R_{\mathrm{p}c}$--$\log S_{\mathrm{H}\alpha }$ relation. For PN K 1-9, the CS candidate (\textit{Gaia} DR3
3053413132594418688) has a high reliability score and a distance consistent with \citetalias{deng26}, but its \(T_{\rm
eff} = 4959\)~K from \(\textit{Gaia}\) photometry makes it an unlikely CSPN. A similar issue occurs for KFL 2 and
PHRJ0905-5548, where the \(\textit{Gaia}\) sources near the geometric centers have \(T_{\mathrm{eff}}\) values of 4948~K
and 5036~K, respectively. Finally, while we identified the CS of H 2-39, its parallax is negative and the distance
estimated by \citetalias{deng26} probably relies on the prior used by \citet{bailer-jones21}, whereas our distance is
estimated using the prior $\log R_{\mathrm{p}c}$--$\log S_{\mathrm{H}\alpha }$.

\subsection{The effect of extinction \label{appendix_c1}} Galactic extinction represents another potential source of
discrepancies in distance estimates. Extinction within the Galaxy varies with longitude and, while $A_V \approx 2$ mag
is typical for most of the bulge, it can reach values as high as 50 mag in regions near the Galactic Center
\citep{nataf13}. Furthermore, extinction towards the bulge is characterized not only by high values but also by a
non-standard extinction law \citep{nataf13, cavichia17}.

\begin{figure*}[!ht] \centering \includegraphics[width=1.0\linewidth]{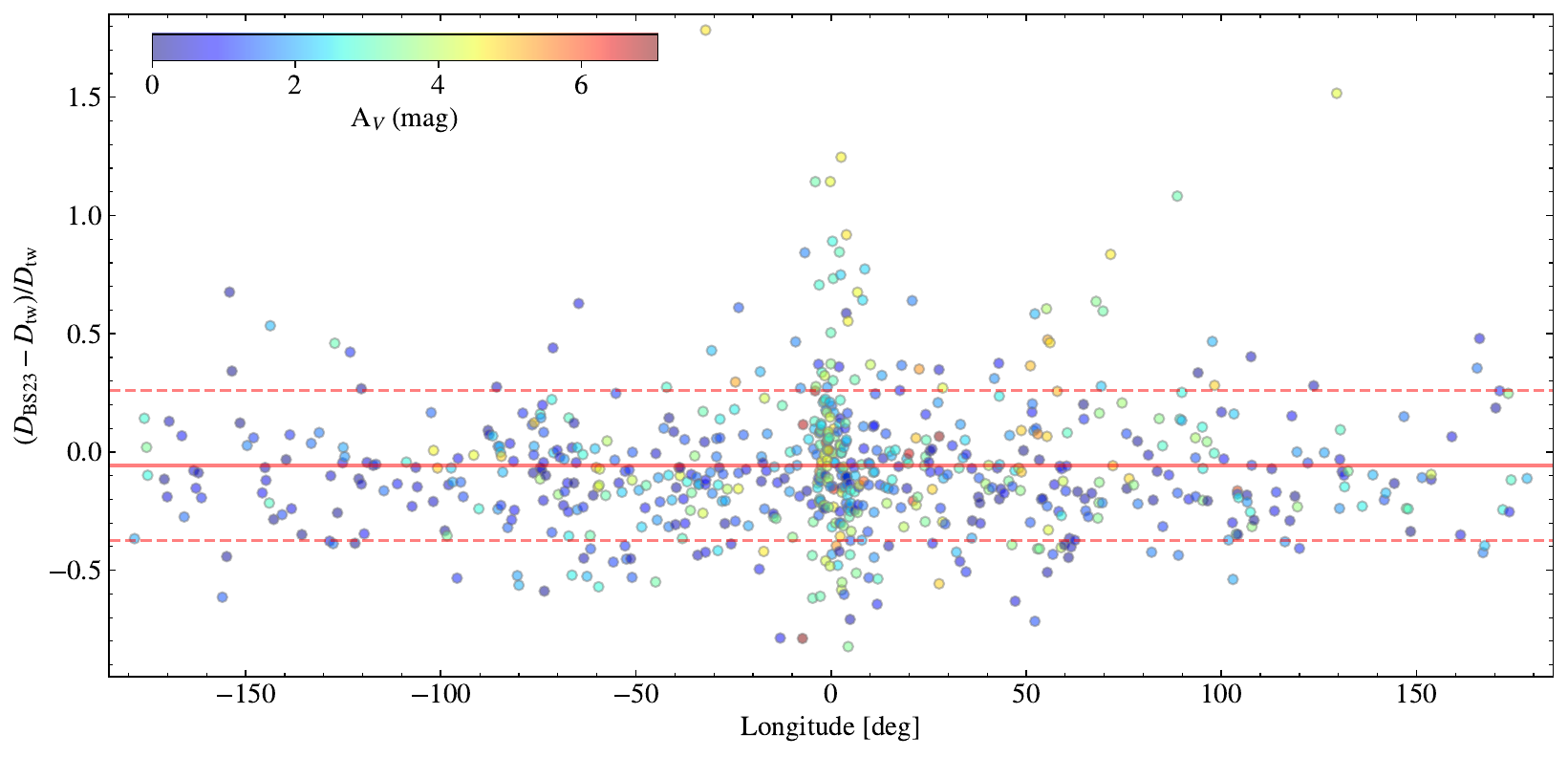}

\caption{Relative error of the distances from \citetalias{bucciarelli23} ($D_{\mathrm{BS23}}$) and the final distances
derived in this work ($D_{\mathrm{tw}}$) as a function of Galactic longitude. The horizontal red continuous line
represents the average relative error and the red dashed lines indicate the standard deviation. The color bar shows the
interstellar extinction ($A_V$) obtained as described in Section \ref{sec:data}.  \label{fig:dists_comparision_av}}
\end{figure*}

The impact of extinction on distances derived using statistical scales based on $S_{\rm H\beta}$
\citepalias{bucciarelli23} and $S_{\rm H\alpha}$ (this work) is illustrated in Fig. \ref{fig:dists_comparision_av},
which plots the relative distance error as a function of Galactic longitude. Relative distance errors increase in
regions towards the Galactic bulge, where extinction is higher and the H$\beta$ fluxes are more severely affected than
H$\alpha$ fluxes. Moreover, the H$\alpha$ flux is typically at least three times brighter than the H$\beta$ flux,
however, for high-extinction Galactic PNe it can be a hundred times or more brighter \citep{cavichia17}, allowing for
high signal-to-noise ratios even for distant Galactic PNe subject to high extinction.

Fig. \ref{fig:dists_comparision_av} also reveals an asymmetry in the relative distance errors: distances from
\citetalias{bucciarelli23} tend to be larger for high-extinction PNe, suggesting an underestimation of extinction.
Consequently, accurate interstellar extinction correction is crucial for obtaining reliable distances for these objects.
For high-extinction PNe, statistical distances based on H${\alpha}$ fluxes are therefore preferred over those based on
H${\beta}$ fluxes.

\newpage

\bibliography{biblio} \bibliographystyle{aasjournal}



\end{document}